\documentclass[12pt]{article}
\usepackage{amsfonts,amssymb,epsfig,amsmath,mathtools}
\usepackage{color}

\usepackage{mathrsfs}
\usepackage{graphicx}
\usepackage{subcaption}

\renewcommand{\baselinestretch}{1.2}

\begin{document}

\makeatletter \@addtoreset{equation}{section} \makeatother
\renewcommand{\theequation}{\thesection.\arabic{equation}}
\renewcommand{\thefootnote}{\alph{footnote}}

\begin{titlepage}

\begin{center}
\hfill {\tt KIAS-P22048}\\
\hfill {\tt SNUTP22-001}\\

\vspace{2cm}

{\Large\bf From giant gravitons to black holes}

\vspace{2cm}

\renewcommand{\thefootnote}{\alph{footnote}}

{\large Sunjin Choi$^1$, Seok Kim$^2$, Eunwoo Lee$^2$ and Jehyun Lee$^2$}

\vspace{0.7cm}

\textit{$^1$School of Physics, Korea Institute for Advanced Study,
Seoul 02455, Korea.}\\

\vspace{0.2cm}

\textit{$^2$Department of Physics and Astronomy \& Center for
Theoretical Physics,\\
Seoul National University, Seoul 08826, Korea.}\\

\vspace{0.7cm}

E-mails: {\tt sunjinchoi@kias.re.kr, seokkimseok@gmail.com\\
eunwoo42@snu.ac.kr, ljs9125@snu.ac.kr}

\end{center}

\vspace{1cm}

\begin{abstract}

We study AdS$_5$ black holes from a recently suggested giant graviton expansion
formula for the index of $U(N)$ maximal super-Yang-Mills theory.
We compute the large $N$ entropy at fixed charges and giant graviton numbers $n_I$
by a saddle point analysis, and further maximize it in $n_I$.
This agrees with the dual black
hole entropy in the small black hole limit. To get black holes at general sizes,
one should note that various giant graviton indices cancel because gauge theory
does not suffer from a Hagedorn-like
pathology by an infinite baryonic tower. With one assumption on the mechanism
of this cancellation, we account for the dual black hole entropy at general sizes.
We interpret our results as analytic continuations of the large $N$ free energies
of SCFTs, and based on it compute the entropies of AdS$_{4,7}$ black holes
from M5, M2 giant gravitons.

\end{abstract}

\end{titlepage}

\renewcommand{\thefootnote}{\arabic{footnote}}

\setcounter{footnote}{0}

\renewcommand{\baselinestretch}{1}

\tableofcontents

\renewcommand{\baselinestretch}{1.2}

\section{Introduction}

Understanding the Hilbert space is the most basic task in
quantum physics. It has also been a key problem of large $N$ gauge theories and
AdS/CFT, often with a focus on the emergent bulk descriptions. For instance,
consider 4d maximal super-Yang-Mills theory on $S^3\times\mathbb{R}$,
dual to the type IIB theory
on global $AdS_5\times S^5$. We normalize the energy $E$ to be dimensionless by
multiplying the AdS radius. At $E\sim N^0$, the spectrum is described by
the gas of gravitons.
At $E\sim \sqrt{Ng_{\rm YM}^2}$, where $g_{\rm YM}$ is the Yang-Mills coupling,
stringy excitations enter. At $E\sim N^1$, a novel finite $N$
effect enters. On the QFT side, this comes from the finite size of the $N\times N$
matrices, imposing trace relations on gauge-invariant operators.
In the gravity dual, this is realized as the gravitons polarizing to
D3-branes. The branes can stretch either in $S^5$, called giant
gravitons \cite{McGreevy:2000cw}, or in $AdS_5$, called dual giant gravitons
\cite{Grisaru:2000zn,Hashimoto:2000zp}. The trace relations are
realized either by giant gravitons having a maximal size, or the
dual giant gravitons having a maximal number
\cite{Bena:2004qv,Suryanarayana:2004ig}. These descriptions
use probe D-brane approaches, whose validity requires that the energy
is not too large, say $E\ll N^2$. This is merely a technical
limitation, and the concept of giant gravitons may exist at
higher energies and provide useful insights. This turned out to be the case
in the half-BPS sector \cite{Lin:2004nb}.

At $E\sim N^2$, semiclassical black hole solutions represent ensembles
of states. In this paper, we wish to clarify the giant graviton picture
at $E\sim N^2$, studying how black holes emerge from a
giant graviton description of the spectral problem.
We consider the BPS sector of the maximal super-Yang-Mills theory
through the index of \cite{Romelsberger:2005eg,Kinney:2005ej}.
This index has been studied to better understand the dual BPS black holes
\cite{Gutowski:2004ez}:
see \cite{Cabo-Bizet:2018ehj,Choi:2018hmj,Benini:2018ywd} and references thereof.
We shall study the recently suggested reformulation of this index
\cite{Imamura:2021ytr,Gaiotto:2021xce,Murthy:2022ien,Lee:2022vig,Imamura:2022aua},
called `giant graviton expansion.' We shall mainly consider the formula by
Yosuke Imamura \cite{Imamura:2021ytr}.\footnote{We understand that
there is a subtlety in the contour choices in this formula, related to
a chemical potential called $a_{\rm loop}$ \cite{Imamura:2021ytr,Lee:2022vig}.
We shall comment on it in section 2 when it seems to be relevant.}

The giant graviton expansion of the index is given by
\begin{equation}\label{index-gg}
  Z(\Delta_I,\omega_i)=Z_{\rm KK}(\Delta_I,\omega_i)
  \sum_{n_1,n_2,n_3=0}^\infty e^{-N\sum_{I=1}^3n_I\Delta_I}
  Z_{n_1,n_2,n_3}(\Delta_I,\omega_i)\ \ ,\ \ I=1,2,3\ ,\ i=1,2\ .
\end{equation}
See section 2 for detailed explanations.
$n_I$'s are winding numbers of maximal giant gravitons along three different
$S^3$ cycles in $S^5$. $Z_{n_1,n_2,n_3}$ is `formally' an index of a
$U(n_1)\times U(n_2)\times U(n_3)$ quiver gauge theory, consisting of
4d/2d fields on the D-branes or at their intersections.
When $E\sim N^2$, we expect all $n_I$'s are typically at the order of $N^1$.
Our strategy of studying this index is roughly as follows.
We first find certain large $n_I$ ($\sim N^1$) saddle points of the integral
representation of $Z_{n_1,n_2,n_3}$. The contour of this integral is complicated and
empirically determined only for low $n_I$'s. The contour information is in principle
important to decide whether a saddle is relevant for approximating the
integral, through the Picard-Lefschetz theory. As often done in practical studies of
challenging integrals, we ignore this issue and assume that our saddles are relevant.
After Legendre transforming the free energy at large fixed charges $q_I$ and $n_I$'s,
one obtains a macroscopic entropy $S(q_I,n_I)$. Further maximizing it in $n_I$'s to
find the dominant term $Z_{n_1,n_2,n_3}$, one would naively find the entropy at fixed
$q_I$. For a reason to be explained, this strategy is correct only in the
`small black hole limit' $\frac{q_I}{N^2}\ll 1$.

To understand why (\ref{index-gg}) is a subtle formula, one should note that
$n_I$'s are the numbers of determinant operators in gauge theory, which are
morally baryons. Baryons and mesons provide towers of confining spectrum,
responsible for fast growth of the high energy density of states.
However, since the basic degrees of freedom are gluons,
the growth for gauge theories should be much slower. Therefore,
for an expansion like (\ref{index-gg}) to correctly capture the gauge
theory entropy at high energies, one expects substantial cancellations
of different $Z_{n_1,n_2,n_3}$'s.
For instance, if such cancellations do not happen, we shall see that the
series (\ref{index-gg}) exhibits very fast growth at large $n_I$'s
and cause a Hagedorn-like pathology \cite{Hagedorn:1965st,Atick:1988si}:
due to string and brane states, the canonical partition function becomes ill-defined.
This implies that individual $S(q_I,n_I)$ and $Z_{n_1,n_2,n_3}$ lose
physical meanings at high energies, while the series (\ref{index-gg}) itself
may be physical after cancellations. To address the cancellations rigorously, one should
be able to compute the subleading terms in the large $N$ limit. This is beyond the scope
of this paper (and the subleading terms depend on the contour choice for
$Z_{n_1,n_2,n_3}$). We shall rather assume a particular mechanism of how the apparently
leading contributions $e^{S(q_I,n_I)}$ cancel, and then proceed to compute the true
entropy that exactly accounts for the dual black holes. As we explain in section 2.2, the
mechanism we suggest assumes that the summations over discrete $n_I$'s can be approximated by integrations of these variables, which is valid only if the subleading terms in the
large $N$ limit are arranged suitably.

As a byproduct, we find an emergent 2d QFT-like structure in our large $N$
calculation when the three chemical potentials for the $U(1)^3\subset SO(6)$ electric
charges are equal. This has to do with the integrand of $Z_{n_1,n_2,n_3}$ reducing
to the ratios of theta functions. This
concretely justifies a study of \cite{Kinney:2005ej}, which assumed
the existence of a hypothetical 2d CFT on the worldvolume of giant gravitons and then
proceeded to count small black holes.

Our large $N$ results can be interpreted as an analytic
continuation  of the maximal super-Yang-Mills index, extending
the idea of \cite{Gaiotto:2021xce,Imamura:2022aua}.
After establishing this interpretation on $AdS_5\times S^5$, we apply it to the
large $N$ index on $AdS_4\times S^7$ based on the expansion of M5-brane giant
gravitons \cite{Arai:2020uwd}. Namely, just assuming the existence of such
an expansion and very basic structures, we explain the entropy of the
AdS$_4$ black holes from the large $N$ free energy of 6d SCFTs on M5-branes.
Similarly, we find a relation between the entropies of black holes
on $AdS_7\times S^4$ and the large $N$ free energies of
3d SCFTs on M2-branes.

The rest of this paper is organized as follows. In section 2.1,
we present a saddle point analysis of $Z_{n_1,n_2,n_3}$, and show that
it accounts for the small black hole entropy. In section 2.2
we explain a possible way in which different $Z_{n_1,n_2,n_3}$ can cancel
at general charges. Then assuming this, we compute the true asymptotic large
$N$ entropy accounting for the dual black holes. Section 2.3 comments on the
similar analysis with three unequal electric charges.
In section 3, we make an interpretation of our results from analytic continuations
and generalize it to account for the entropies of BPS black holes in
AdS$_{4,7}$. Section 4 concludes with discussions.

\section{Giant graviton index and black holes}

The index for the $\mathcal{N}=4$ Yang-Mills theory is defined by
\begin{equation}
  Z(\Delta_I,\omega_i)={\rm Tr}\left[(-1)^Fe^{-\sum_{I=1}^3\Delta_I Q_I
  -\sum_{i=1}^2\omega_iJ_i}\right]
\end{equation}
subject to the condition $\sum_{I=1}^3\Delta_I-\sum_{i=1}^2\omega_i=2\pi i\mathbb{Z}$,
where $Q_I$ are $U(1)^3\subset SO(6)$ R-charges and $J_i$ are
$U(1)^2\subset SO(4)$ angular momenta. See \cite{Romelsberger:2005eg,Kinney:2005ej}
for a unitary matrix integral representation of this index for the $U(N)$ gauge group.
Recently, an alternative expression for this index was proposed.
It takes the form of (\ref{index-gg}), where
$Z_{\rm KK}$ is the index of low energy gravitons \cite{Kinney:2005ej}.
$Z_{n_1,n_2,n_3}(\Delta_I,\omega_i)$ is given by a
$U(n_1)\times U(n_2)\times U(n_3)$
matrix integral of the form \cite{Imamura:2021ytr}
\begin{equation}
  Z_{n_1,n_2,n_3}=\oint \prod_{I=1}^3\prod_{a=1}^{n_I}du^{(I)}_{a}
  \cdot \prod_{I=1}^3Z^{\rm 4d}_I\cdot Z^{\rm 2d}_{I,I+1}\ \ ,\ \ \
  \textrm{where }\ I+3\sim I\ ,
\end{equation}
while $Z_{0,0,0}\equiv 1$.
The functions $Z^{\rm 4d}_I$ and $Z^{\rm 2d}_{I,I+1}$ appearing in the integrand
are given as follows. From now on, let us define
$\Delta_I\equiv -2\pi i\tau_I=-2\pi i(\tau+z_I)$ with $\sum_{I=1}^3z_I=0$,
and $\omega_1=-2\pi i(\frac{3\tau}{2}+y-1)$,
$\omega_2=-2\pi i(\frac{3\tau}{2}-y)$.
Then $Z^{\rm 4d}_3$ from the 4d $U(n_3)$ adjoint fields
is given by
\begin{equation}
  Z^{\rm 4d}_3=\prod_{a,b}\frac{\Gamma(u_{ab}^{(3)}-\tau-z_3;\tau+z_1,\tau+z_2)
  \Gamma(u_{ab}^{(3)}+\frac{3}{2}\tau\pm y;\tau+z_1,\tau+z_2)}
  {\Gamma(u_{ab}^{(3)};\tau+z_1,\tau+z_2)}\ .
\end{equation}
Here and below, whenever the argument contains $\pm$, corresponding two functions
are multiplied. $u^{(I)}_{ab}\equiv u^{(I)}_a-u^{(I)}_b$,
and $\Gamma(z;\sigma,\tau)$ is the elliptic Gamma function defined by
\begin{equation}
  \Gamma(z;\sigma,\tau)=\prod_{m,n=0}^\infty
  \frac{1-e^{-2\pi iz}e^{2\pi i((m+1)\sigma+(n+1)\tau)}}
  {1-e^{2\pi iz}e^{2\pi i(m\sigma+n\tau)}}\ .
\end{equation}
Other $Z^{\rm 4d}_I$ are given similarly
by permuting the $I=1,2,3$ indices. The integrand from the 2d
$U(n_1)\times U(n_2)$ bifundamental fields is given by
\begin{equation}
  Z^{\rm 2d}_{1,2}=\prod_{a=1}^{n_1}\prod_{b=1}^{n_2}
  \frac{\theta(\pm (u_{ab}^{(12)}+a_{12})+\frac{\tau_3}{2}+y,\tau_3)}
  {\theta(\pm (u_{ab}^{(12)}+a_{12})-\tau+\frac{z_3}{2},\tau_3)}\ ,
\end{equation}
where $u^{(I,I+1)}_{ab}\equiv u^{(I)}_a-u^{(I+1)}_b$. $\theta(z,\tau)$ is
the $q$-theta function (with `$q$' given by $e^{2\pi i\tau}$) defined by
\begin{equation}
  \theta(z,\tau)=\prod_{n=0}^\infty(1-e^{2\pi iz}e^{2\pi in\tau})
  (1-e^{-2\pi iz}e^{2\pi i(n+1)\tau})\ .
\end{equation}
Other $Z^{\rm 2d}_{I,I+1}$ are given similarly. The integration contour is
complicated, and is related to how the auxiliary parameters $a_{I,I+1}$ are
chosen. Only the value of $a_{\rm loop}\equiv a_{12}+a_{23}+a_{31}$ is important.
In \cite{Imamura:2021ytr} and
\cite{Lee:2022vig}, two different choices of $a_{\rm loop}$ were made, also
with different choices of the integration contour.
Both prescriptions are tested till certain low orders.
One of $a_{\rm loop}=-\frac{3\tau}{2}\pm y$ was chosen in \cite{Imamura:2021ytr},
while $a_{\rm loop}=0$ was chosen in \cite{Lee:2022vig}.
The situation might be that both prescriptions work to all orders
in $n_I$'s, or one of the two is correct for higher $n_I$'s.
Although we have little to say about this issue, we simply note that our
saddle point ansatz below works with the choice $a_{\rm loop}=0$
of \cite{Lee:2022vig}. Perhaps with the choice of \cite{Imamura:2021ytr},
residue contributions may be more important when the contour crosses
poles during its deformation towards the saddle point.
(Such an issue may also arise in the original Yang-Mills matrix integral for the index,
as commented on in \cite{Choi:2021rxi}.) So we set
$a_{I,I+1}=0$ from now on.

We would like to study the large $N$ behaviors of $Z_{n_1,n_2,n_3}$. Since $n_I$'s
contribute $Nn_I$ to the electric charges $Q_I$, which we want to
scale as $N^2$, we let $n_I$'s to scale linearly in $N$.

\subsection{Large $N$ saddle points and small black holes}

In this subsection we shall consider the index at $\Delta_1=\Delta_2=\Delta_3$.
(We shall comment on the generalization to unequal $\Delta_I$'s in section 2.3.)
This corresponds to taking the $z_I\rightarrow 0$ limit. It was shown
\cite{Imamura:2021ytr} that individual $Z_{n_1,n_2,n_3}$ diverges in this limit,
while the full index after summing them over remains finite. We are interested in
the leading order free energy $\log Z_{n_1,n_2,n_3}\sim N^2$ in this limit.
To understand this limit more precisely, we first decompose the contributions to
$Z^{\rm 4d}_I$ from the $N$ Cartans at $a=b$ and the off-diagonals at $a\neq b$.
The former part can be written in the limit as
\begin{equation}
  \left[\frac{-e^{2\pi i\tau}\theta(\frac{\tau}{2}-y,\tau)}
  {(1-e^{2\pi iz_{I+1,I}})(1-e^{2\pi iz_{I-1,I}})E(\tau)^2}\right]^{n_I}\ ,
\end{equation}
where $z_{I,J}\equiv z_I-z_J$, $E(\tau)\equiv\prod_{n=1}^\infty(1-e^{2\pi in\tau})$.
So the Cartan parts diverge in the limit $z_I\rightarrow 0$. This accounts for
many of the divergences encountered in \cite{Imamura:2021ytr} in this limit.
The divergence is linear in $N$, $\sim N\log \varepsilon$ at small $z_I\sim\varepsilon$.
So we ignore this part since we are interested in the leading free
energy proportional to $N^2$. (However, see the later part of this subsection and
section 2.2 for important roles of the subleading parts.)
The off-diagonal part with $a\neq b$ contains extra divergences in the limit
$z_I\rightarrow 0$, by the $z_I$ dependent poles pinching the integration
contour \cite{Imamura:2021ytr}. Our precise setting of taking the limit is as
follows. We are interested in the behaviors of the integrand near the saddle point
of our interest, to be presented below. The saddle point is away from the contour,
and will not suffer in any sense from the pinching of the $z_I$ dependent poles.
So as for this part, we naively take the $z_I\rightarrow 0$ limit and simplify
the integrand. Using $\Gamma(z+\sigma;\sigma,\tau)=\theta(z,\tau)\Gamma(z;\sigma,\tau)$,
$\Gamma(z+\tau;\sigma,\tau)=\theta(z,\sigma)\Gamma(z;\sigma,\tau)$ and
$\Gamma(z;\sigma,\tau)=\frac{1}{\Gamma(\sigma+\tau-z;\sigma,\tau)}$,
one obtains
\begin{equation}\label{Z-4d-theta}
  Z^{\rm 4d}_I\stackrel{z_I\rightarrow 0}{\longrightarrow}
  \prod_{1\leq a\neq b\leq n_I}\frac{\theta(u^{(I)}_{ab}+\frac{\tau}{2}-y,\tau)}
  {\theta(u^{(I)}_{ab}-\tau,\tau)}\ .
\end{equation}
We have ignored the Cartan part which only makes a subleading
$N^1$ contribution.  At the saddle point, all $u_a^{(I)}$'s will be different,
so that this function remains finite in the $z_I\rightarrow 0$ limit.
We realize that
the contributions from the 4d fields are given in terms of the theta
functions after substantial cancellations. Similarly,
$Z^{\rm 2d}_{I,I+1}$ are given in the $z_I\rightarrow 0$ limit by
\begin{equation}\label{Z-2d-theta}
  Z^{\rm 2d}_{I,I+1}=\prod_{a=1}^{n_I}\prod_{b=1}^{n_{I+1}}
  \frac{\theta(\pm u_{ab}^{(I,I+1)}+\frac{\tau}{2}-y,\tau)}
  {\theta(\pm u_{ab}^{(I,I+1)}-\tau,\tau)}\ .
\end{equation}
$Z^{\rm 4d}_{I}$ and $Z^{\rm 2d}_{I,I+1}$ in this limit are
invariant under shifting $u_a^{(I)}$ to $u_a^{(I)}+1$ or $u_a^{(I)}+\tau$.

As our large $N$ (and large $n_I\sim N$)
saddle point ansatz, we take each set of $U(n_I)$ eigenvalues $u^{(I)}_a$ to be
uniformly distributed along the $\tau$-circle,
\begin{equation}\label{ansatz-2d}
  u^{(I)}=x_I\tau\ ,\ \ 0<x_I<1\ ,\ \ \rho(x_I)=1\ .
\end{equation}
This is a coarse-grained continuum description of the
eigenvalues, which are separated from their nearest neighbor by a distance at order
$\frac{1}{N}$. We can typically assume that none of these eigenvalues are at precisely
the same values. Therefore, (\ref{Z-4d-theta}) and (\ref{Z-2d-theta}) do not diverge
due to $u_a^{(I)}=u_b^{(J)}$. If such a divergence apparently seems to happen in the
continuum description, it should be avoided by integrating over
$x_I$'s with a principal-value prescription.
It is easy to see that this distribution solves the large $n_I$ saddle point
equation. To check this, it is convenient to first S-dualize the integrand
using the identity
\begin{equation}
  \theta(z,\tau)=e^{-\pi iB(z,\tau)}
  \theta({\textstyle \frac{z}{\tau}},{\textstyle -\frac{1}{\tau}})\ \ ,\ \ \
  B(z,\tau)\equiv {\textstyle \frac{z^2}{\tau}+z\left(\frac{1}{\tau}-1\right)
  +\frac{1}{6}\left(\tau+\frac{1}{\tau}\right)-\frac{1}{2}}\ .
\end{equation}
We shall set $y$ to be in the range $0<y_1<1$, where
$y\equiv y_1+y_2\tau$ with real $y_1,y_2$: this convention
can be chosen by a suitable period shift of $y$. Then, applying the S-dual
identities, one obtains
\begin{eqnarray}
  Z^{\rm 4d}_I&\sim&
  \exp\left[\pi in_I^2{\textstyle \left(
  \frac{y-y^2}{\tau}-\frac{3}{2}+\frac{9\tau}{4}\right)}\right]
  \prod_{1\leq a\neq b\leq n_I}
  \frac{\theta(\frac{u_{ab}^{(I)}-y}{\tau}+\frac{1}{2},-\frac{1}{\tau})}
  {\theta(\frac{u_{ab}^{(I)}}{\tau},-\frac{1}{\tau})}\\
  Z^{\rm 2d}_{I,I+1}&=&\exp\left[2\pi in_In_{I+1}{\textstyle\left(
  \frac{y-y^2}{\tau}-\frac{3}{2}+\frac{9\tau}{4}
  \right)}\right]\prod_{a=1}^{n_I}\prod_{b=1}^{n_{I+1}}
  \frac{\theta(\frac{\pm u_{ab}^{I,I+1}-y}{\tau}+\frac{1}{2},-\frac{1}{\tau})}
  {\theta(\frac{\pm u_{ab}^{I,I+1}}{\tau},-\frac{1}{\tau})}\ .\nonumber
\end{eqnarray}
Collecting all, the integrand is given by
a constant factor
\begin{equation}\label{s-dual-pert}
  \exp\left[\frac{\pi i(n_1+n_2+n_3)^2
  \left(y-\frac{3\tau}{2}\right)\left(1-y-\frac{3\tau}{2}\right)}{\tau}
  \right]
\end{equation}
times
\begin{equation}\label{s-dual-nonpert}
  \tilde{Z}(u^{(I)})=\prod_{I=1}^3\left[\tilde{Z}_I^{\rm 4d}\tilde{Z}_{I,I+1}^{\rm 2d}
  \right]\equiv\prod_{I=1}^3\left[\prod_{1\leq a\neq b\leq n_I}
  \frac{\theta(\frac{u_{ab}^{(I)}-y}{\tau}+\frac{1}{2},-\frac{1}{\tau})}
  {\theta(\frac{u_{ab}^{(I)}}{\tau},-\frac{1}{\tau})}\cdot
  \prod_{a=1}^{n_I}\prod_{b=1}^{n_{I+1}}
  \frac{\theta(\frac{\pm u_{ab}^{I,I+1}-y}{\tau}+\frac{1}{2},-\frac{1}{\tau})}
  {\theta(\frac{\pm u_{ab}^{I,I+1}}{\tau},-\frac{1}{\tau})}\right]\ .
\end{equation}
In order to show that (\ref{ansatz-2d}) is a saddle point, one should show that the
force $\frac{\partial}{\partial u_a^{(I)}}\log\tilde{Z}$ vanishes in the large $N$ limit.
More precisely, one should show that the leading $N^1$ order term of the force vanishes.
This force is given by
\begin{equation}\label{force}
  -\frac{N}{\tau}\int_0^1 dx^\prime\frac{\partial}{\partial x^\prime}
  \left[\log \tilde{Z}^{\rm 4d}_I(u(x)\!-\!u(x^\prime))
  +\log \tilde{Z}_{I,I+1}^{\rm 2d}(u(x)\!-\!u(x^\prime))+
  \log \tilde{Z}^{\rm 2d}_{I-1,I}(u(x^\prime)\!-\!u(x))\right]\ .
\end{equation}
The expression inside the square bracket of the right hand side is given by
a linear combination of the function of the form
$\log(1-e^{\pm 2\pi i(x-x^\prime)}e^{-\frac{2\pi i(n+\alpha)}{\tau}})$
with $n\in\mathbb{Z}\geq 0$, $\alpha\geq 0$ and $x,x^\prime\in[0,1]$.
($\alpha$ may be either $0$ or $y$.) So all these log functions are periodic in
$x^\prime\rightarrow x^\prime+1$ shift without crossing the branch cut. Therefore,
we integrate the derivative of a periodic function over a circle,
which vanishes. The terms with $\alpha=0$ and $n=0$ have the branch points
on the circle $x^\prime\in [0,1]$, but employing the principal-valued integrals
as explained, they also vanish.

One can compute $\log Z_{n_1,n_2,n_3}$ at this saddle point. By evaluating
$\log \tilde{Z}$ in the continuum limit, similar to the evaluation of
(\ref{force}), one finds $\log \tilde{Z}=0$. This is because
the integral of $\log(1-e^{\pm 2\pi i(x-x^\prime)}e^{-\frac{2\pi i(n+\alpha)}{\tau}})$
is zero at $n\geq 0$, $\alpha\geq 0$.
So the large $N$ free energy is given by
\begin{equation}\label{log-Z-cardy}
  \log Z_{n_1,n_2,n_3}=\frac{\pi in^2
  \left(y-\frac{3\tau}{2}\right)\left(1-y-\frac{3\tau}{2}\right)}{\tau}\ ,
\end{equation}
where $n\equiv n_1+n_2+n_3$. Note that the leading free energy depends only
on one combination of $n_I$. So there
apparently is a large number of degenerate terms if we only consider
the leading free energy.
One can Legendre transform this free energy to obtain the macroscopic
entropy at fixed charges $q$, $j$ conjugate to $\tau$, $y$, also at fixed
$n$. This amounts to the extremization of
\begin{equation}
  S(q,j;\tau,y,n)=\frac{\pi in^2 y(1-y)}{\tau}-\frac{3\pi in^2}{2}
  +\frac{9\pi in^2}{4}\tau-2\pi i\tau\cdot (3q-nN)
  -2\pi i y\cdot j\ .
\end{equation}
The charges correspond to $q=\frac{Q_1+Q_2+Q_3}{3}+\frac{J_1+J_2}{2}$,
$j=J_1-J_2$.
The solution is given by
\begin{equation}\label{saddle-legendre}
  \tau=i\frac{n^2 }{2\sqrt{2n^2P-j^2}}\ ,\ \
  y=\frac{1}{2}-i\frac{j}{2\sqrt{2n^2P-j^2}}
\end{equation}
where $P\equiv 3q-nN-\frac{9n^2}{8}$, and the extremized entropy is given by
\begin{equation}\label{entropy-2d}
  S(q,j,n)=\frac{\pi}{2}
  \sqrt{n^2(24q-8nN-9n^2)-4j^2}-\pi ij-\frac{3\pi in^2}{2}\ .
\end{equation}
The constant imaginary term $-\pi ij$ can be ignored in the discussions below.

Before proceeding, let us comment on the structure of the asymptotic
$Z_{n_1,n_2,n_3}$ that we obtained in (\ref{log-Z-cardy}).
We first investigate the
structure of the expansion (\ref{index-gg}) in the grand canonical ensemble with
fixed $\tau$, $y$. Since these parameters are complex, it is helpful to
focus on a region which contains the saddle point of the Legendre
transformation (\ref{saddle-legendre}).\footnote{See
\cite{Choi:2018vbz,Agarwal:2020zwm,Choi:2021lbk}.
Basically, the phases of fugacities should be tuned in the index even
in the grand canonical ensemble, to minimize the unwanted boson-fermion
`cancellations' during macroscopic approximations.}
For instance, as for $\tau$, let us take it to be purely imaginary with
${\rm Im}(\tau)>0$. For $y$, let us freeze $y=\frac{1}{2}$ for simplicity of
the discussion. This corresponds to setting $j=0$ in the microcanonical ensemble,
or unrefining the chemical potential $y$ for $j$ in the grand canonical ensemble.
Then one finds that the giant graviton expansion (\ref{index-gg}) takes the form of
\begin{equation}\label{index-gg-large-n}
  Z\sim \!\sum_{n_1,n_2,n_3}\!\Omega(n_I)e^{2\pi iN\tau n}
  \exp\left[\frac{\pi in^2}{4\tau}(1-3\tau)^2\right]=
  \!\sum_{n_1,n_2,n_3}\!\Omega(n_I)e^{-Nn\beta}\exp\left[
  \frac{\pi^2 n^2}{2}
  \left(\frac{1}{\beta}-\frac{3i}{\pi}-\frac{9\beta}{4\pi^2}
  \right)\right]
\end{equation}
where $\beta\equiv-2\pi i\tau$ is real and positive. Recall that
$n\equiv n_1+n_2+n_3$, and $\Omega(n_I)$ come from
the subleading contributions to $\log Z_{n_1,n_2,n_3}$
in the large $n\sim N$ expansion.
At large $n$, each $|Z_{n_1,n_2,n_3}|$ grows very fast like $e^{an^2}$
with certain $a>0$ when $\beta<\beta_c\equiv\frac{2\pi}{3}$.
So at high temperatures, unless the subleading factors $\Omega(n_I)$ are given in a
manner that various terms substantially cancel,
the sum will diverge very badly at large $n$.
One can be more realistic and insert the complex values of
$\tau(q)$ as a function of real charge $q$, at which we know that
BPS black hole saddle points exist \cite{Choi:2018hmj,Choi:2018vbz}.
Then one finds that ${\rm Im}\left[\frac{1}{\tau(q)}-9\tau(q)\right]<0$ is always
met, again making $|Z_{n_1,n_2,n_3}|$ to grow fast.
However, if the expansion (\ref{index-gg}) provides an exact expression for the gauge
theory partition function, we expect the series (\ref{index-gg-large-n}) to better
behave at high temperatures where the system deconfines
\cite{Choi:2018vbz,Copetti:2020dil,Choi:2021lbk}.

Let us elaborate more on why we expect the series (\ref{index-gg-large-n})
to behave well for $\beta<\beta_c$. For instance, consider a series
of the form
\begin{equation}\label{series-hagedorn}
  \sum_n \Omega(n)e^{\beta_c n}e^{-n\beta}
\end{equation}
with $\beta_c>0$, and $\Omega(n)$ does not affect the exponential growth of
$e^{\beta_c n}$. The series ceases to converge at $\beta<\beta_c$
outside its radius of convergence. If this series is for a thermal partition
function, the divergence is the Hagedorn pathology \cite{Hagedorn:1965st}
caused by exponential growth of the density of states at high energy.
It happens due to an infinite tower of mesonic states \cite{Hagedorn:1965st},
or an infinite tower of string oscillations \cite{Atick:1988si}.
The apparent divergence $\sim e^{a n^2}$ of (\ref{index-gg-large-n}) would make
the series worse-behaved than (\ref{series-hagedorn}).
Namely, unless cancellations happen, the radius of convergence is zero.
One may interpret this divergence (if present) as coming from a much faster
asymptotic growth of baryonic states. However, the notion
of baryonic states should become ambiguous at large charge $q$ for which
$\tau(q)$ enters the deconfining regime. For expressions like
(\ref{index-gg}) or (\ref{index-gg-large-n}) to remain relevant,
$\Omega(n_I)$'s should be arranged so that the apparent asymptotic growth
$e^{an^2}$ cancels. The cancellation effects should be more crucial for
$\tau(q)$ with larger $q$, as the system is deeper inside the deconfining regime.
If such cancellations are not taken into account, each term in the series
(\ref{entropy-2d}) may over-estimate the microcanonical entropy.

With this caution in mind, let us try to extract the microcanonical entropy from
the formula (\ref{entropy-2d}). We first study the case with $j=0$.
Since $n$ is not a physical charge, we should try
to maximize ${\rm Re}[S(q,n)]$ as a function of non-negative integer $n$.
${\rm Re}[S(q,n)]$ is positive when
\begin{equation}
  0<n<\frac{4}{9}\left[-N+\sqrt{N^2+{\textstyle \frac{27q}{2}}}\right]\equiv n_\ast(q)\ .
\end{equation}
So to study the macroscopic entropy from this
index, one only needs to sum over $n$ till $n_\ast(q)$. Thus, we consider
\begin{equation}\label{microcanonical-sum}
  e^{S(q)}=\oint d\tau e^{-2\pi i\tau\cdot 3q}
  \!\!\sum_{n_1,n_2,n_3}\!\!e^{2\pi iNn\tau} Z_{n_1,n_2.n_3}
  \sim \!\!\sum_{n_1,n_2,n_3}^{n\leq n_\ast(q)}\!\!\Omega(n_I)\exp\left[
  \frac{\pi n}{2}\sqrt{24 q-8nN-9n^2}-\frac{3\pi in^2}{2}\right]\ .
\end{equation}
We study this quantity at large $N$ and large $q\propto N^2$,
naively expecting at this moment that the leading contribution comes from certain
$n$ at order $N^1$. We first note that the overall phase factor
$e^{-\frac{3\pi in^2}{2}}$ oscillates between $i$ and $1$, depending on whether
$n$ is odd or even. So dividing the sum into even/odd $n$'s and
naturally expecting that the maximization will not be sensitive to the even/odd
nature of $n$'s, this phase does not matter and the dominant contribution to this
entropy is given by the maximum of ${\rm Re}[S(q,n)]$. This happens at
\begin{equation}
  n_0=\frac{-N+\sqrt{N^2+12q}}{3}\ ,
\end{equation}
which is in the range $0<n_0(q)<n_\ast(q)$. The maximal entropy is given by
\begin{equation}
  {\rm Re}[S(q)]={\rm Re}[S(q,n_0(q))]=
  \frac{\pi(-N+\sqrt{N^2+12 q})}{3\sqrt{6}}\sqrt{N^2+18 q-N\sqrt{N^2+12q}}\ .
\end{equation}
We first study its asymptotic
behaviors at $q\ll N^2$ and $q\gg N^2$, which would respectively correspond to
the small and large black hole limits. One finds
\begin{eqnarray}\label{entropy-overestimate}
  {\rm Re}[S(q)]&=&\frac{\pi(2q)^{\frac{3}{2}}}{N}
  -\frac{9\pi q^{\frac{5}{2}}}{\sqrt{2} N^3}+
  \frac{351\pi q^{\frac{7}{2}}}{8\sqrt{2}N^5}
  -\frac{8937\pi q^{\frac{9}{2}}}{32\sqrt{2}N^7}+
  \frac{1048059\pi q^{\frac{11}{2}}}{512\sqrt{2}N^9}+\cdots\ \
  \textrm{for}\ q\ll N^2 \nonumber\\
  {\rm Re}[S(q)]&=&2\pi q-\frac{4\pi Nq^{\frac{1}{2}}}{3\sqrt{3}}
  +\cdots\ \ \textrm{for}\ q\gg N^2\ .
\end{eqnarray}
One can show that this ${\rm Re}[S(q)]$ is asymptotically equal to the
Bekenstein-Hawking entropy $S_{\rm BH}(q)$ of the dual black hole
when $q\ll N^2$, in which case
${\rm Re}[S(q)]\approx \frac{\pi(2q)^{\frac{3}{2}}}{N}\approx S_{\rm BH}(q)$.
Away from the asymptotic limit $q\ll N^2$, ${\rm Re}[S(q)]>S_{\rm BH}(q)$
always holds. In particular, in the two asymptotic limits,
$S_{\rm BH}$ is expanded as
\begin{eqnarray}\label{entropy-bh}
  S_{\rm BH}(q)&=&\frac{\pi(2q)^{\frac{3}{2}}}{N}
  -\frac{21\pi q^{\frac{5}{2}}}{\sqrt{2}N^3}+\frac{1287 \pi q^{\frac{7}{2}}}
  {8\sqrt{2}N^5}-\frac{46189\pi q^{\frac{9}{2}}}{32\sqrt{2}N^7}
  +\frac{7243275\pi q^{\frac{11}{2}}}{512\sqrt{2}N^9}+\cdots\ \
  \textrm{for}\ q\ll N^2\nonumber\\
  S_{\rm BH}(q)&=&\sqrt{3}\pi\left(\frac{N^2q^2}{2}\right)^{\frac{1}{3}}+\cdots
  \ \ \textrm{for}\ q\gg N^2\ .
\end{eqnarray}
One finds a small over-estimating deviation
${\rm Re}[S(q)]-S_{\rm BH}(q)\approx\frac{12\pi q^{\frac{5}{2}}}{\sqrt{2}N^3}>0$
in the small charge expansion, and ${\rm Re}[S(q)]\gg S_{\rm BH}(q)$ in
the large charge expansion. As already explained, our interpretation of this
over-estimate is that we have been ignoring the possible cancellations
of the apparently leading order terms due to nontrivial $\Omega(n_I)$'s in
(\ref{index-gg-large-n}). The large charge behavior
${\rm Re}[S(q)]\sim 2\pi q=3\beta_c q$ of (\ref{entropy-overestimate})
is a Hagedorn growth.

The agreement of the leading entropy
${\rm Re}[S(q)]\approx \frac{\pi(2q)^{\frac{3}{2}}}{N}$
with the Bekenstein-Hawking entropy of small black holes
might still look a bit miraculous.
To better appreciate this, it is first worthwhile to note that small black
holes are never dominant saddles in the grand canonical ensemble. Also, they
always stay in the confining region in the complex $\tau$ space \cite{Choi:2021lbk}.
So it makes sense that they admit a description in terms of D-branes, which are baryonic
objects in the confining phase. On the other hand, as $q$ gradually grows,
giant gravitons will eventually lose their meaning at high energy. This is because the
fundamental high energy degrees of freedom are gluons rather than their bound states.
Interestingly, the D3-brane giant graviton approach has been already employed in
\cite{Kinney:2005ej} to account for the entropy of small black holes. The calculation
we did with $S(q,n)$ was discussed in \cite{Kinney:2005ej}, in precisely the same
computational procedure. The rough idea of \cite{Kinney:2005ej} is to regard the
maximal giant gravitons to be similar to the wrapped D-branes
which account for 5d asymptotically flat black holes \cite{Strominger:1996sh}.
Since most of the microscopic accounts for asymptotically flat black holes use
branes, and since small AdS$_5$ black holes are (at least mathematically) identical
to the 5d asymptotically flat black holes embedded in large AdS, it is natural
that both objects admit similar D-brane-based descriptions.
We find that our studies provide precise logical grounds for the calculations 
of \cite{Kinney:2005ej}.

From (\ref{saddle-legendre}), the saddle point value of $\tau$ for the Legendre
transformation at $j=0$ is
\begin{equation}
  \tau(q,n)=\frac{in}{\sqrt{24 q-8nN-9n^2}}=\frac{in}
  {3\sqrt{(n_\ast(q)-n)(n+n_\ast(q)+\frac{8N}{9})}}\ .
\end{equation}
In the small black hole limit $q\ll N^2$, $n$ is ranged in
$0<n<n_\ast(q)\approx \frac{3q}{N}$ and the maximum $n_0(q)$ of $S(q,n)$
is approximately $n_0(q)\approx \frac{2q}{N}$. Around the maximum $n_0(q)$,
$\tau$ scales like
\begin{equation}
  |\tau|\sim \frac{n}{\sqrt{q}}\sim \frac{q^{\frac{1}{2}}}{N}\ll 1\ .
\end{equation}
So the small black hole limit $q\ll N^2$ corresponds to the
`Cardy limit' $\tau\rightarrow i0$ in the 2d-like integrand (\ref{Z-4d-theta}),
(\ref{Z-2d-theta}). We find this to be a concrete realization of the studies
made in \cite{Kinney:2005ej}, which assumed the existence of a hypothetical
2d CFT living on the worldvolume of maximal giant gravitons at fixed
$n=n_1+n_2+n_3$ and used its Cardy formula to account for the small black hole
entropy. The 2d CFT was supposed to live on the Hopf fiber circle of
the $S^5$, which is wrapped by the D3-branes.
However, there was no logical justification for the existence of such a 2d
CFT, since the 4d worldvolume has no scale separation which justifies the 2d
reduction. We found from our index in the limit $z_I\rightarrow 0$ that the 4d
part of the integrand $Z^{\rm 4d}_{I}$ partly canceled to
yield Jacobi theta functions, which are 2 dimensional objects.
So what justifies the 2d reduction here is the boson-fermion cancellations
in the index. This is much more specific than a reduction based on the
scale separation. This 2d description may break down if one
studies unprotected quantities beyond
the index. For instance, \cite{Kinney:2005ej} studied the charge relation
satisfied by small black holes, $J_1+J_2\sim \frac{q^2}{N^2}$. The idea of
\cite{Kinney:2005ej} is as follows. If the 2d CFT exists,
small AdS$_5$ black holes are dual to its NS sector. It is related to the CFT in the
Ramond sector by a spectral flow. The Ramond sector CFT describes
5 dimensional asymptotically flat black holes \cite{Strominger:1996sh,Breckenridge:1996is} satisfying a charge relation $J_1+J_2=0$. So if the 2d description exists
universally beyond the index, the spectral flow will connect the charge
relation $J_1+J_2=0$ to that of the AdS black holes. The relation obtained from
this route does not agree with the charge relation of AdS black holes
\cite{Kinney:2005ej}. We interpret this as the absence of the 2d description
beyond the index.

We can generalize the studies to the case with
$j\equiv J_1-J_2\neq 0$ by keeping $y\neq\frac{1}{2}$. Again we are only
able to successfully count small black holes by keeping the leading term
\begin{equation}
  \log Z_{n_1,n_2,n_3}\sim\frac{\pi in^2y(1-y)}{\tau}
\end{equation}
of (\ref{log-Z-cardy}) at small $\tau$.
Making a Legendre transformation of this free energy by extremizing
\begin{equation}
  S(q,j;n)\sim \frac{\pi i n^2y(1-y)}{\tau}-2\pi i\tau(3q-nN)-2\pi i y j
\end{equation}
and then maximizing ${\rm Re}[S(q,j;n)]$ with $n$, one obtains the entropy given by
\begin{equation}
  S(q,j)=\pi\sqrt{\frac{8q^3}{N^2}-j^2}\ .
\end{equation}
$\tau$, $y$ at these saddles always satisfy $|\tau|\ll 1$, ${\rm Re}(y)=\frac{1}{2}$
as long as the charges are away from the closed timelike curve (CTC) bound
$j^2=\frac{8q^3}{N^2}$, staying within the regime that we assumed.
This is precisely the entropy of small spinning BPS black holes in
$AdS_5\times S^5$. This can also be regarded as the BMPV black holes
\cite{Breckenridge:1996is} embedded in large AdS$_5$.
We emphasize that this is the first microscopic counting of small spinning AdS$_5$
black holes at all allowed values of $J_1\neq J_2$. In fact
accounting for black holes at $J_1\neq J_2$ has been technically tricky
from the QFT dual. For instance, in the saddle point approach
to the Yang-Mills matrix model, general $J_1\neq J_2$ was discussed only
in the 4d Cardy limit \cite{Choi:2018hmj}. At general finite charges,
\cite{Choi:2021rxi} found saddle points which cover substantial charge regions
for $j$, but failed to cover the whole parameter space of CTC-free black
holes. More precisely, \cite{Choi:2021rxi} found the saddles when certain
inequalities were met, like eqn.(2.41) or (2.45) there. In the parametrization
of BPS black holes given by the third reference of \cite{Gutowski:2004ez},
these inequalities cover the region $0<ag,bg<1$. On the other hand, the CTC-free
black holes exist in a bigger region $ag,bg<1$, $a+b+abg>0$.
This should be due to our limited understanding of the large $N$
matrix model saddles. At least in the small black hole limit, it is amusing that
the giant graviton calculation of this paragraph was able to cover
the whole CTC-free black holes satisfying $ag, bg\ll 1$,
$a+b>0$.

\subsection{Comments on finite size black holes}

From the absence of the Hagedorn behavior in gauge theories,
we think it is obvious that cancellations of different $Z_{n_1,n_2,n_3}$'s
happen in general. However, computing such cancellations at large $N$ is technically very
challenging. This is because the cancellations
happen due to relative minus signs, whose precise determination goes
beyond the leading order calculation.
For instance, we tried to compute such subleading terms in the small $\tau$ regime,
but found that the precise integral contours for $Z_{n_1,n_2,n_3}$ are needed to compute them. Also, with such a contour dependence, taking
$z_I\rightarrow 0$ limit is trickier than in the previous subsection.

In this subsection, leaving the full microscopic analysis to the future,
we shall make a simple assumption on how these subleading corrections should be arranged.
This assumption will allow us to compute the true entropy of this index after the
cancellations, which precisely reproduces the dual black hole entropy.
The claim is that, once we include the  $\frac{1}{N}$ effects to each
$Z_{n_1,n_2,n_3}$'s contribution, the degeneracy at given $n$ will be lifted
by small deviations from (\ref{log-Z-cardy}) in a way that the sum over $n$ can
be replaced by an integral. The microcanonical sum
(\ref{microcanonical-sum}) over discrete $n$ can be replaced by
\begin{equation}\label{sum-integral}
  e^{S(q)}=\oint d\tau e^{-2\pi i\tau\cdot 3q}
  \!\!\sum_{n_1,n_2,n_3}\!\!e^{2\pi iNn\tau} Z_{n_1,n_2.n_3}
  \sim \int_0^{n_\ast(q)}\! dn\exp\left[
  \frac{\pi n}{2}\sqrt{24 q-8nN-9n^2}-\frac{3\pi in^2}{2}\right]\ .
\end{equation}
The claim asserts that we use the same function $S(q,n)$ but sum over a dense set
of $n$'s. Before explaining anything about this claim, we emphasize that we have
no derivation of (\ref{sum-integral}) except that this formula will give the exact
black hole entropy at an arbitrary size.

Let us first explain why this is a nontrivial claim, and in particular why
it is related to including $\frac{1}{n_I}$ subleading terms.
Our claim is essentially that, once
we include the subleading corrections at fixed $n$,
$Z_{n_1,n_2,n_3}$ will behave like
\begin{equation}\label{dense}
  Z_{n_1,n_2,n_3}\sim \exp\left[S(q,n+\delta_n(n_I))\right]\ ,\ \
  S(q,x)\equiv\frac{\pi x}{2}\sqrt{24q-8Nx-9x^2}-\frac{3\pi ix^2}{2}
\end{equation}
with a nontrivial function $\delta_n(n_I)\sim\mathcal{O}(1)\ll n$ of $n_I$.
This makes the distribution of $n+\delta_n(n_I)$ dense over a range of
$\mathcal{O}(1)$ width around $n$.
(\ref{dense}) is a claim about the $\frac{1}{N}$ subleading corrections of
$Z_{n_1,n_2,n_3}$, since all $\delta_n(n_I)$ dependent terms are subleading.
If this happens to all values of $n$, one would obtain
\begin{equation}
  \sum_{n=0}^{n_\ast(q)}Z_{n_1,n_2,n_3}=\int_0^{n_\ast(q)}
  \rho(n)\exp\left[S(q,n)\right]
\end{equation}
where $\rho(n)$ is a suitable distribution determined by
$\delta_n(n_I)$'s. Since the total number of summands
satisfying $n\leq n_\ast(q)$ is proportional to $n_\ast^3$, one finds
$\int_0^{n_\ast(q)} dn\rho(n)\sim n_\ast^3\sim N^3$. So
$\log\rho(n)$ is a logarithmic correction
to $S(q,n)\sim N^2$. Thus, ignoring it, one obtains (\ref{sum-integral}).

If the sum over $n$ is replaced by an integral over $n$, it is no longer valid to
find the dominant contribution by maximizing ${\rm Re}[S(q,n)]$. Rather, one should
find a saddle point of $S(q,n)$ in the complex $n$ plane. Regarding $S(q,n)$ as a
complex function of $n$, the maxima $n_0(q)$ of ${\rm Re}[S(q,n)]$ on the real axis
is not a saddle point, due to nontrivial ${\rm Im}[S(q,n)]=-\frac{3\pi n^2}{2}$
for continuous $n$. In summary, part of our claim is about the $\frac{1}{n_I}$ subleading
corrections of ${\rm Im}[S(q,n)]$, which lift the degeneracy and render substantial
cancellations of different $Z_{n_1,n_2,n_3}$'s.

Given (\ref{sum-integral}), one can
identify the saddle point on the complex $n$ plane. In fact it is
inconvenient to work directly with the last expression of (\ref{sum-integral}).
Rather, we keep the variables $\tau,y$ unintegrated, and consider the multiple
integral formula for $e^{S(q,j)}$ given by
\begin{eqnarray}\label{3d-integral}
  e^{S(q,j)}&\sim&\oint d\tau \oint dy \int_0^{n_\ast(q)}dn
  e^{\log Z_{n_1,n_2,n_3}-2\pi i\tau(3q-nN)-2\pi iy\cdot j}\\
  &\sim& \oint d\tau \oint dy \int_0^{n_\ast(q)}dn~
  \exp\left[\frac{\pi in^2(y-\frac{3\tau}{2})(1-y-\frac{3\tau}{2})}{\tau}
  -2\pi i\tau(3q-nN)-2\pi iy\cdot j\right]\nonumber\ .
\end{eqnarray}
Note that at this stage we reintroduced the refinement with $y$ or $j$, since
the analysis is no more difficult. To find the possible saddle points in the complex $n$
plane, we can simply extremize the 3-dimensional integral (\ref{3d-integral}).
(Later in this subsection, when numerically discussing the contour deformation,
it will be more convenient to use the original
1 dimensional integral (\ref{sum-integral}).)
It is easy to first extremize in $n$, since the integrand is Gaussian
in $n$. One finds the saddle point
$n_{\rm s}=-\frac{N\tau^2}{(y-\frac{3\tau}{2})(1-y-\frac{3\tau}{2})}$. Inserting
this, the remaining $\tau,y$ integral is given by
\begin{equation}
  S^{S(q,j)}\sim\oint d\tau \oint dy \exp\left[
  -\frac{\pi i N^2\tau^3}{(\frac{3\tau}{2}-y)(\frac{3\tau}{2}-1+y)}
  -2\pi i\tau\cdot 3q-2\pi iy\cdot j\right]\ .
\end{equation}
Reintroducing  $\omega_1=-2\pi i\left(\frac{3\tau}{2}+y-1\right)$,
$\omega_2=-2\pi i\left(\frac{3\tau}{2}-y\right)$,
the exponent is given by
\begin{equation}
  \frac{N^2}{2}\frac{\left(\frac{\omega_1+\omega_2}{3}-\frac{2\pi i}{3}\right)^3}
  {\omega_1\omega_2}+\omega_1(Q+J_1)+\omega_2(Q+J_2)
\end{equation}
where $q\equiv Q+\frac{J_1+J_2}{2}$, $j\equiv J_1-J_2$.
This is precisely the entropy function (at equal electric charges $Q_I$) for
the Bekenstein-Hawking entropy of BPS black holes
in $AdS_5\times S^5$ \cite{Hosseini:2017mds}. Although we discussed the
saddle points of the 3-dimensional integral (\ref{3d-integral}), the same entropy
is obtained with $S(q,n)$ from the last expression of (\ref{sum-integral}).

\begin{figure}[!t]
\centering
\includegraphics[width=0.45\textwidth]{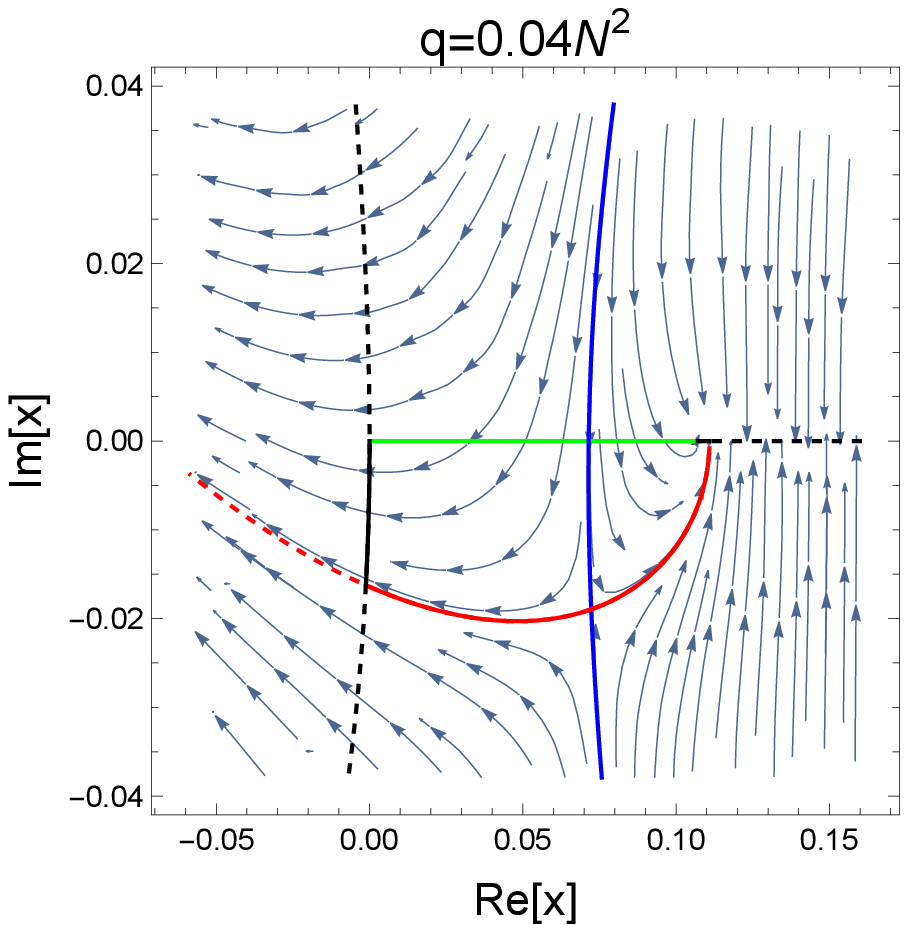}
\hspace{.3cm}\includegraphics[width=0.45\textwidth]{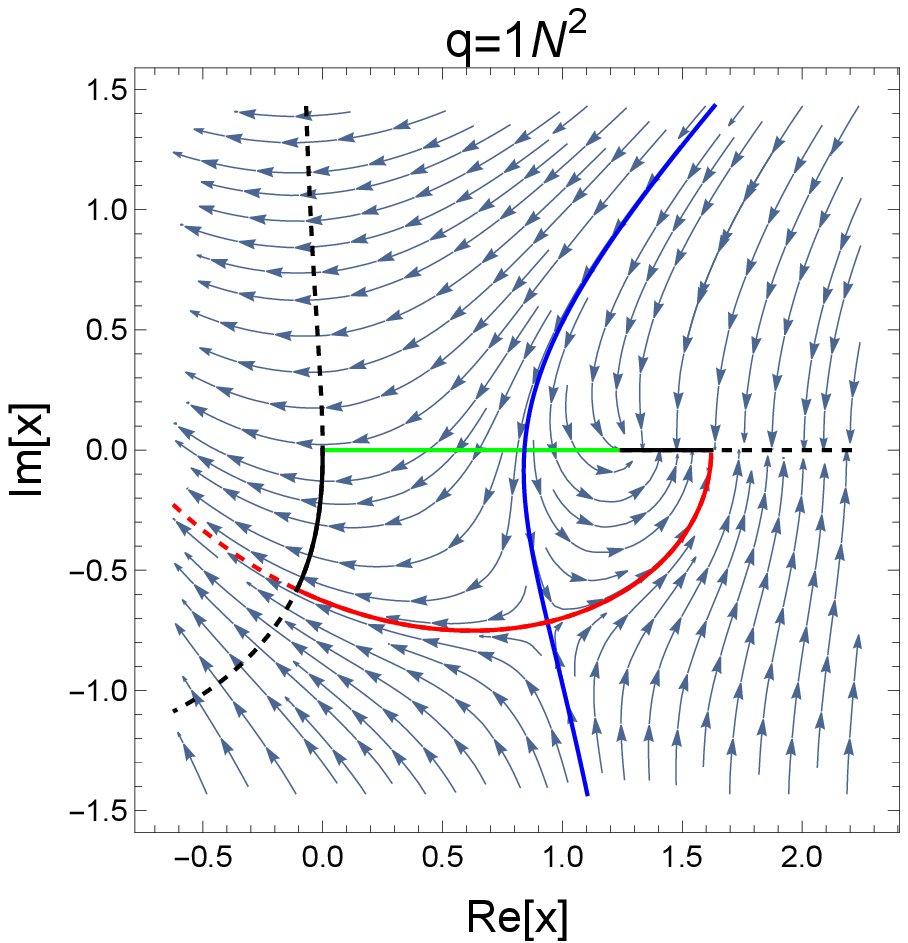}\\
\includegraphics[width=0.45\textwidth]{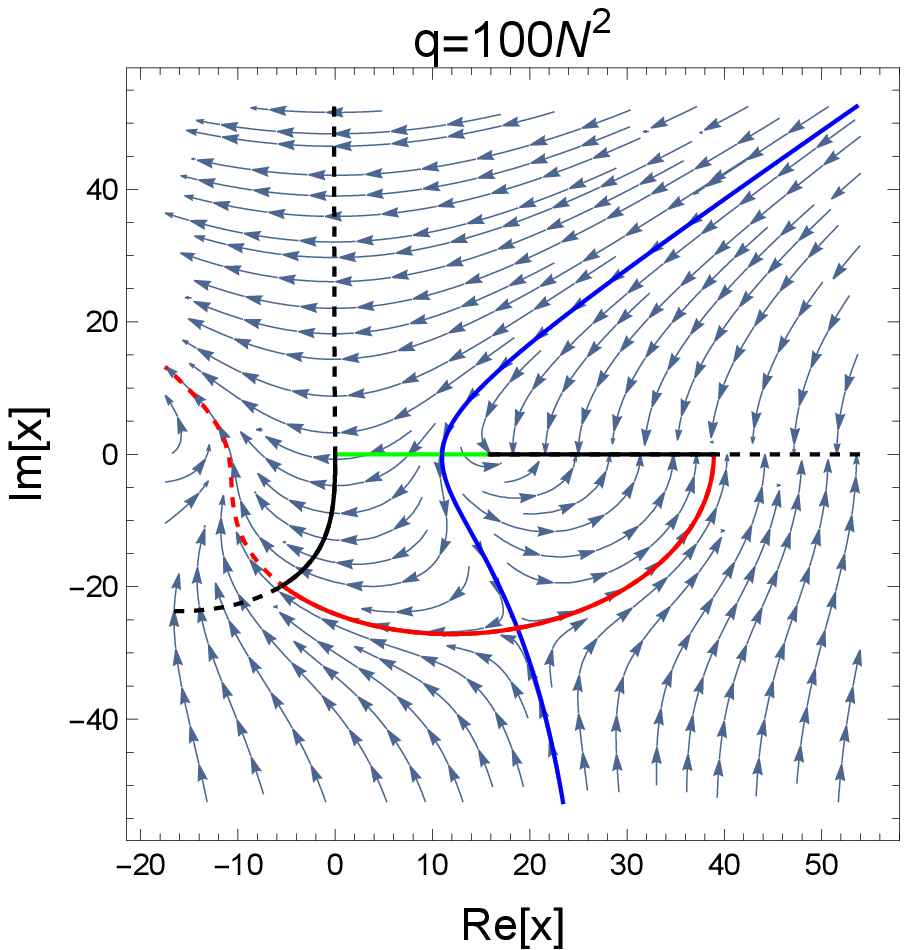}
\hspace{.3cm}\includegraphics[width=0.45\textwidth]{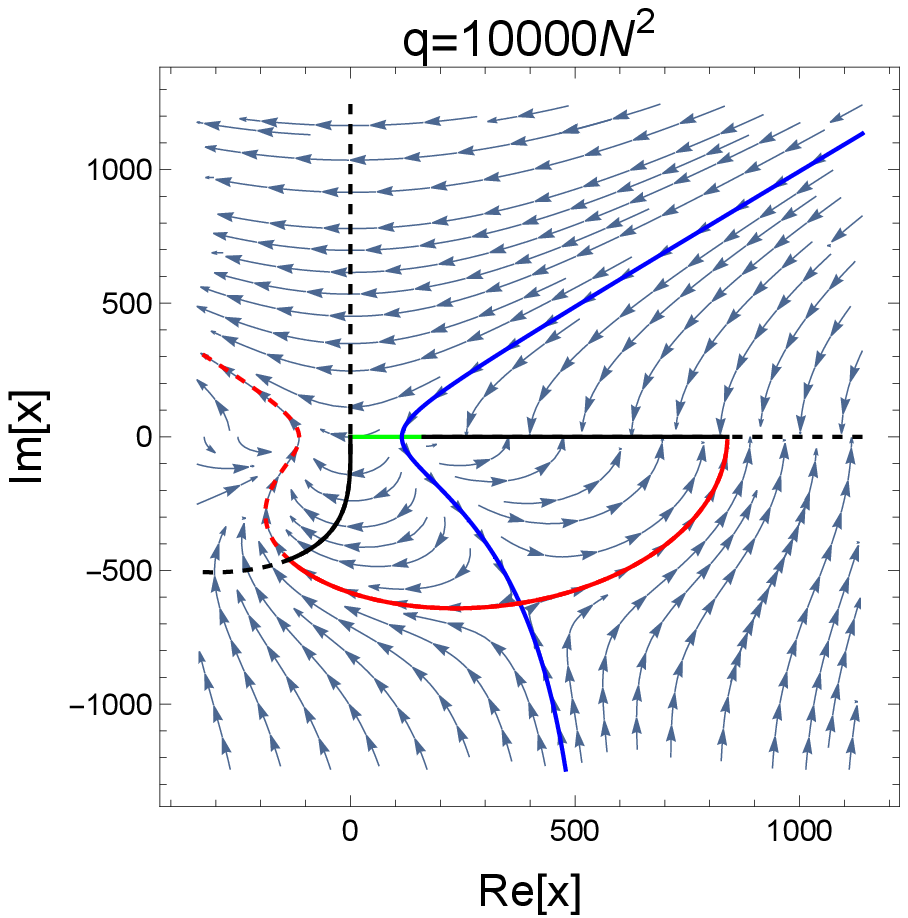}
\caption{Contour deformations at various $q$.
Green interval is the original integration contour $[0,x_\ast]$.
Blue arrows denote the gradient flow which determines the
contour deformation. The deformed contour is the union of two solid black lines
$C_1$, $C_3$, and the solid red line $C_2$. The solid blue line is the
steepest ascent contour.}
\label{picard-lefschetz}
\end{figure}

We finally show that the contour can be deformed to pass through this saddle,
by extending the standard Picard-Lefschetz theory. Consider the following integral
\begin{equation}
  \int_0^{x_\ast(\frac{q}{N^2})}dx e^{N^2 f(x)}\ \ ,\ \
  f(x)\equiv \frac{\pi x}{2}\sqrt{24{\textstyle \frac{q}{N^2}} - 8 x-9x^2}
  -\frac{3\pi ix^2}{2}\ ,
\end{equation}
where $x=\frac{n}{N}$ and
$x_\ast({\textstyle \frac{q}{N^2}})=\frac{n_\ast(q)}{N}$.
If the integrand vanishes at the two ends $x=0,x_\ast$, one can deform the integration
contour to the steepest descent contour. The steepest descent contour has maximal
${\rm Re}[f(x)]$ at the saddle point, and satisfies the stationary phase condition
${\rm Im}[f(x)]=$ constant. The integration on this contour can be approximated at large
$N$ by a Gaussian approximation around the saddle point. In our case, the integrand does
not vanish at the two ends. Then the steepest descent contour passing through our complex
saddle point $x_{\rm s}\equiv \frac{n_{\rm s}}{N}$ does not end on $x=0,x_\ast$,
so we have to slightly extend this standard method.  We combine an interval $C_2$ of
the steepest descent contour (solid red line of Fig. \ref{picard-lefschetz}) with two more intervals $C_1$, $C_2$ of contours satisfying ${\rm Re}[f]=$ constant and ending on
$x=0,x_\ast$, respectively (solid black). The original contour
$C_0=[0,x_\ast]$ (green) can be deformed to $C_1\cup C_2\cup C_3$.
As $q$ decreases, one can see that the complex saddle $x_{\rm s}$ approaches
the real maximum $x_0\equiv \frac{n_0(q)}{N}$ of ${\rm Re}[S(q,n)]$.
So the complex
saddle approach naturally converges the naive analysis with real
$n$ of section 2.1 in the small black hole limit.

The dominant term of the integral on $C_2$ can be computed
by the Gaussian approximation around $x=x_{\rm s}$, yielding a
term of the form $\sim e^{N^2f(x_{\rm s})}$. Then, denoting the two ends of
the interval $C_2$ by $x_1$ and $x_3$, respectively, the integrands on
$C_1$ and $C_3$ take the form of
\begin{equation}
  e^{N^2 {\rm Re}[f(x_1)]}\int_{C_1} dx e^{iN^2 {\rm Im}[f(x)]}\ ,\ \
  e^{N^2 {\rm Re}[f(x_3)]}\int_{C_3} dx e^{iN^2 {\rm Im}[f(x)]}\ ,
\end{equation}
respectively. Since $C_2$ is the steepest descent contour, one
finds $e^{N^2 {\rm Re}[f(x_{1,3})]}\ll e^{N^2 {\rm Re}[f(x_{\rm s})]}$
and these integrals are bounded as
\begin{equation}
  \left|e^{N^2 {\rm Re}[f(x_{1,3})]}\int_{C_{1,3}} dx e^{iN^2 {\rm Im}[f(x)]}\right|
  \leq e^{N^2 {\rm Re}[f(x_{1,3})]}\int_{C_{1,3}} dx \left|e^{iN^2 {\rm Im}[f(x)]}\right|
  \ll e^{N^2 {\rm Re}[f(x_{\rm s})]}\ .
\end{equation}
Therefore, the contribution from $C_1\cup C_3$ is subdominant, justifying the
approximation using the Gaussian approximation near $x_{\rm s}$. As illustrated
in Fig. \ref{picard-lefschetz}, we checked for a wide range of $\frac{q}{N^2}$ that
the contour can always be deformed in this way.

\subsection{Comments on unequal electric charges}

So far, we studied the index with the chemical potentials $\Delta_I$ for
the three electric charges $Q_I$ unrefined,
$\Delta_1=\Delta_2=\Delta_3\equiv -2\pi i\tau$. In this subsection we comment
on the generalizations with unequal $\Delta_I$'s. Note that in the original
Yang-Mills matrix model of \cite{Romelsberger:2005eg,Kinney:2005ej}, taking
independent $\Delta_I$ was rather straightforward,
while introducing the refinement $y\neq \frac{1}{2}$ for two independent angular momenta
$J_1,J_2$ was much trickier \cite{Choi:2021lbk,Choi:2021rxi}. This was basically
because independent $\omega_1,\omega_2$ for the spacetime charges in QFT could
introduce branch points in the matrix model potentials which yield nonzero eigenvalue
forces. In fact, as explained in section 2.1,
\cite{Choi:2021rxi} found saddle points for the black
holes at $J_1\neq J_2$ only when certain inequalities are met: see
eqn. (2.41) or (2.45) of \cite{Choi:2021rxi}.
In the giant graviton index $Z_{n_1,n_2,n_3}$, since the role of internal and
spacetime symmetries are partly exchanged, such as
$\Delta_1,\Delta_2\leftrightarrow \omega_1,\omega_2$ for the giant gravitons with
$n_3\neq 0$, the situation is the other way round. We have seen in section 2.1
that the giant graviton approach sees the black holes at $J_1\neq J_2$ (i.e. at
$y\neq\frac{1}{2}$) rather easily at least when the black hole size is small.
This is because $\omega_{1,2}$ are the chemical potentials for internal symmetries
in this approach. On the other hand, we find it very difficult to
construct saddle points of $Z_{n_1,n_2,n_3}$ when $\Delta_I$'s are different.

More concretely, we have tried to construct the large $n_I$ saddle points by
noting that the integrand $Z^{\rm 4d}_I$ resembles the integrand of a Yang-Mills
index with $U(n_I)$ gauge group, except for the tachyon and contour issues
\cite{Imamura:2021ytr,Lee:2022vig}.
So we tried to use the ansatz of \cite{Choi:2021rxi} to find the
saddle point at unequal $\Delta_I$'s. This almost solves the saddle point
equations but not quite, due to several branch points in the potential.
During the course, however, we could write down a free energy for
$\log Z_{n_1,n_2,n_3}$ whose Legendre transformation and maximization in $n_I$'s
yield the dual black hole entropy. Although we have
a gap in our derivation, we strongly believe that we found the correct
answer. So we simply report our findings without any microscopic derivation.

We find that the refined free energy should be given by
\begin{equation}\label{log-Z-unequal}
  \log Z_{n_1,n_2,n_3}=\frac{\pi i(n_1\tau_1+n_2\tau_2+n_3\tau_3)^2}{\tau_1\tau_2\tau_3}
  \left({\textstyle \frac{1}{2}\sum_{I=1}^3}\tau_I-y\right)
  \left({\textstyle \frac{1}{2}\sum_{I=1}^3}\tau_I+y-1\right)
\end{equation}
and the corresponding entropy function to extremize is
\begin{equation}\label{entropy-ftn-unequal}
  S(q_I,j;n_I,\tau_I,y)=\log Z_{n_1,n_2,n_3}-2\pi i\sum_I\tau_I (q_I-n_IN)
  -2\pi iy\cdot j\ .
\end{equation}
Note that $q_I\equiv Q_I+\frac{J_1+J_2}{2}$, $j\equiv J_1-J_2$.
Like the analysis we did for equal $\tau_I$,
we should first extremize this in $\tau_I$'s, and then maximize the real part
with $n_I$. If we do this calculation, again one generally obtains an entropy
${\rm Re}[S(q_I,j)]$ which overestimates the degeneracy unless cancellations
of various $Z_{n_1,n_2,n_3}$ are taken into account.

Like the case with equal $q_I$, the entropy estimated with
single $Z_{n_1,n_2,n_3}$ reproduces the black hole entropy in the small black hole limit
$q_I,j\ll N^2$. For simplicity, we show this only at $j=0$ ($y=\frac{1}{2}$).
The small black hole limit corresponds to $|\tau_I|\ll 1$,
in which case one obtains
\begin{equation}
  S(q_I;n_I,\tau_I)\approx\frac{\pi i(n_1\tau_1+n_2\tau_2+n_2\tau_3)^2}
  {4\tau_1\tau_2\tau_3}-2\pi i\sum_I\tau_I(q_I-n_IN)\ .
\end{equation}
This is a real function for purely imaginary $\tau_I$'s. We shall extremize
$S$ with $\tau_I$'s on this subspace. Since $S$ is real, the next
maximization of ${\rm Re}[S]$ with $n_I$ is just maximizing $S$.
We can exchange the order of the two extremizations. $S(q_I;n_I,\tau_I)$ is quadratic
in $n_I$, and depends only on $n_1\tau_1+n_2\tau_2+n_3\tau_3$.
Therefore, only this linear combination of three $n_I$ is fixed after
extremizing with $n_I$, leaving two parameters unfixed. This is a generalization
of sections 2.1 and 2.2 where only $n_1+n_2+n_3\equiv n$ was fixed.
After this extremization, one obtains
\begin{equation}
  S(q_I;\tau_I)\approx-4\pi iN^2\tau_1\tau_2\tau_3-2\pi i\sum_I\tau_Iq_I\ .
\end{equation}
This is precisely the entropy function for the small black holes, whose
further extremization yields the Bekenstein-Hawking entropy
$S(q_I)\approx\frac{\pi}{N}\sqrt{8q_1q_2q_3}$ of small AdS black holes.

Beyond small black holes, we also expect that subleading order terms should
render substantial cancellations of different $Z_{n_1,n_2,n_3}$'s, in order
for the index not to exhibit a Hagedorn-like pathology at large $n_I$'s.
At complex $\tau_I$, one can only fix
$\sum_I n_I\tau_I$ so that one of the three parameters $n_I$ remains
unfixed in the leading free energy. Summing over them could render cancellations.
We also suggest that the concrete mechanism of such cancellation is
replacing the sum over discrete $n_I$'s by an integral, as we explained in
section 2.2. This allows one to seek a complex saddle point for
$n_I$'s in (\ref{entropy-ftn-unequal}). Since (\ref{log-Z-unequal}) is
quadratic in $n_I$'s, one first finds a Gaussian saddle point for
$n_1\tau_1+n_2\tau_2+n_3\tau_3$, obtaining
\begin{equation}
  S(q_I,j;\tau,y)=-\frac{\pi iN^2\tau_1\tau_2\tau_3}
  {\left(\frac{1}{2}\sum_I\tau_I-y\right)
  \left(\frac{1}{2}\sum_I\tau_I+y-1\right)}
  -2\pi i\sum_I\tau_I q_I-2\pi i y\cdot j\ .
\end{equation}
This is precisely the entropy function of BPS black holes in $AdS_5\times S^5$
\cite{Hosseini:2017mds}, whose further extremization yields
the Bekenstein-Hawking entropies of the dual black holes.

\section{Analytic continuation and AdS$_{4,7}$ black holes}

In this section, we interpret our results in section 2 as the analytic continuation
of the maximal super-Yang-Mills index. \cite{Gaiotto:2021xce} established such an
interpretation for $Z_{0,0,n_3}$. See also \cite{Imamura:2022aua}. For general
$Z_{n_1,n_2,n_3}$, we find a similar interpretation of its large $N$ free energy.

Let us first review \cite{Gaiotto:2021xce} in the language of \cite{Imamura:2021ytr}.
The integrand for $Z_{0,0,n_3}$ is simply $Z^{\rm 4d}_3$ of our section 2.1,
as the quiver consists only of one adjoint node. This
is related to the integrand $Z^{\rm YM}_{\rm int}$
of 4d maximal super-Yang-Mills index $Z^{\rm YM}_{U(n_3)}$ in a very simple manner.
Let us first note that, when we write the arguments of $Z^{\rm YM}_{\rm int}$ as
$Z^{\rm YM}_{\rm int}(\Delta_1,\Delta_2,\Delta_3;\omega_1,\omega_2;u_a)$, the first three
denote the $U(1)^3\subset SO(6)$ internal rotations while the next two
denote the $U(1)^2\subset SO(4)$ rotations on the spacetime of the QFT.
Then $Z^{\rm 4d}_3$ is given by \cite{Imamura:2021ytr,Gaiotto:2021xce}
\begin{equation}\label{Z3-analytic-continuation}
  Z^{\rm 4d}_3=Z^{\rm YM}_{\rm int}(\omega_1,\omega_2,-\Delta_3;\Delta_1,\Delta_2
  ;u_a^{(3)})\ .
\end{equation}
This formula can be understood as follows. Since $n_3$
D3-branes wrap $S^3\subset S^5$, the two worldvolume rotation parameters
are $\Delta_1,\Delta_2$. On the other hand,
$SO(4)$ rotations on AdS$_5$ are internal symmetries on D3-branes.
So $\omega_1,\omega_2$ are their internal rotation parameters. Finally,
since the maximal giant gravitons can shrink rather than grow, losing energies,
the corresponding transverse scalar is tachyonic. This demands replacing
$\Delta_3$ by $-\Delta_3$. In fact, as emphasized in \cite{Imamura:2021ytr,Lee:2022vig},
(\ref{Z3-analytic-continuation}) has to be defined by analytic
continuation since ${\rm Re}(-\Delta_3)<0$. Suitably choosing the integration
contour, one finds that $Z_{0,0,n_3}$ is obtained from
$Z^{\rm YM}_{U(n_3)}$ by exchanging $\omega_{1,2}\leftrightarrow \Delta_{1,2}$
and replacing $\Delta_3\rightarrow -\Delta_3$ with analytic continuation
\cite{Gaiotto:2021xce}.

One can also get its large $n_3\sim N$ free energy from analytic continuation.
When $\omega_{1,2}$, $-\Delta_3$, $\Delta_{1,2}$ on the right hand side of
(\ref{Z3-analytic-continuation}) have positive real parts, its large $N$ free energy
is given by
\begin{equation}\label{4d-continue}
  \log Z_{0,0,n_3}=\log Z^{\rm YM}_{U(n_3)}
  (\omega_1,\omega_2,-\Delta_3;\Delta_1,\Delta_2)
  \sim \frac{n_3^2}{2}\frac{\omega_1\omega_2(-\Delta_3)}{\Delta_1\Delta_2}
\end{equation}
where the imaginary parts of the chemical potentials are suitably shifted by
their periods to satisfy either
$\omega_1+\omega_2-\Delta_3-\Delta_1-\Delta_2=\pm 2\pi i$. This result can be understood
in two different ways. Firstly, it can be understood as derived from various calculations
of the Yang-Mills index \cite{Choi:2018hmj,Benini:2018ywd,Choi:2021rxi}. Secondly, one can interpret it as the free energy of dual AdS$_5$ black holes
\cite{Hosseini:2017mds,Cabo-Bizet:2018ehj}. For 4d maximal super-Yang-Mills, both
viewpoints are available. Having in mind less explored SCFT's, to be
explored later in this section, we emphasize the virtue of understanding
$\log Z_{\rm SCFT_D}$ as the free energy of dual black holes in AdS$_{D+1}$.
Once (\ref{4d-continue}) is known in the region ${\rm Re}(-\Delta_3)>0$,
it is quite immediate to continue it to the physical region
${\rm Re}(\Delta_3)>0$. Namely, we just keep the expression on the right hand
side of (\ref{4d-continue}). This continuation assumes the absence
of the Stokes' phenomena. Let us assume this and proceed.
In the parametrization of section 2, we take
$\Delta_I=-2\pi i\tau_I$,
$\omega_1=-2\pi i\left(\frac{1}{2}\sum_I\tau_I+y-1\right)$ and
$\omega_2=-2\pi i\left(\frac{1}{2}\sum_I\tau_I-y\right)$ which satisfy
$\Delta_1+\Delta_2+\Delta_3-\omega_1-\omega_2=-2\pi i$.
Then (\ref{4d-continue}) is given by
\begin{equation}
  \log Z_{0,0,n_3}\sim\frac{\pi in_3^2\tau_3
  (\frac{1}{2}\sum_I\tau_I-y)(\frac{1}{2}\sum_I\tau_I+y-1)}{\tau_1\tau_2}\ ,
\end{equation}
which is the giant graviton free energy (\ref{log-Z-unequal})
or (\ref{log-Z-cardy}) at $n_1=n_2=0$.
Therefore, our formulae of section 2 can be
naturally understood as the analytic continuation of the Yang-Mills index.

In fact one can similarly interpret our general formula (\ref{log-Z-unequal})
for $Z_{n_1,n_2,n_3}$. Expanding the complete square in the numerator, one obtains
\begin{eqnarray}\label{log-Z-unequal-factorize}
  \log Z_{n_1,n_2,n_3}&=&\sum_{I=1}^3\frac{\pi in_I^2\tau_I
  (\frac{1}{2}\sum_J\tau_J-y)(\frac{1}{2}\sum_J\tau_J+y-1)}{\tau_{I-1}\tau_{I+1}}\\
  &&+\sum_{I=1}^3\frac{2\pi in_{I-1}n_{I+1}(\frac{1}{2}\sum_J\tau_J-y)
  (\frac{1}{2}\sum_J\tau_J+y-1)}{\tau_I}\nonumber
\end{eqnarray}
where $I\sim I+3$ is understood. The three terms on the first line are
$\log Z_{n_1,0,0}$, $\log Z_{0,n_2,0}$ and $\log Z_{0,0,n_3}$. One can more
concretely identify each of them as the saddle point value of the integrand
$\log Z^{\rm 4d}_I$. This was derived in section 2.1 at equal $\tau_I$'s. At equal
$\tau_I$'s, the three terms on the second line can also be separately identified
as the saddle point values of the integrands $\log Z^{\rm 2d}_{I,I+1}$.
So we naturally find a picture of the large $N$ giant graviton free
energy, as the sum of three maximal super-Yang-Mills free energies
and three 2d free energies at the intersections.

We comment on two features of (\ref{log-Z-unequal}) and
(\ref{log-Z-unequal-factorize}). Firstly, the six
4d and 2d contributions factorize in (\ref{log-Z-unequal-factorize}).
This does not always have to be the case, as there is no reason for these degrees
of freedom to decouple. We find that it is a rather exceptional property, perhaps for
even dimensional QFT's whose free energies can be read off
from anomalies. A more fundamental aspect is the $n_I$
dependence through $\sum_{I=1}^3 n_I\Delta_I$. This may be
heuristically understood as follows.
$\frac{1}{8}$-BPS giant graviton solutions of \cite{Mikhailov:2000ya} are given by
holomorphic surfaces in $\mathbb{C}^3\supset S^5$
\begin{equation}\label{1/8-BPS}
  0=\sum_{n_1,n_2,n_3=0}^\infty C_{n_1,n_2,n_3}z_1^{n_1} z_2^{n_2} z_3^{n_3}
\end{equation}
where $z_I$ are the coordinates of $\mathbb{C}^3$. $\Delta_I$ can be regarded
as $U(1)^3$ rotation parameters on the moduli space, transforming
$z_1^{n_1} z_2^{n_2} z_3^{n_3}\rightarrow e^{\sum_I n_I\Delta_I}
z_1^{n_1} z_2^{n_2} z_3^{n_3}$. The formula of \cite{Imamura:2021ytr} can be
interpreted as an `equivariant localization' of an integration over the
moduli space given by $C_{n_1,n_2,n_3}$'s (modded out by an overall multiplication
of a complex number) \cite{Gaiotto:2021xce,imamura-talk}. When all $\Delta_I$'s assume
general values,
(\ref{1/8-BPS}) is invariant under the rotation only if a single
term is kept on the right hand side. So the moduli are completely lifted.
In this case, $n_I$'s label a discrete set of points in the moduli space which
are invariant under $U(1)^3$.
When $\Delta_I$'s assume rational ratios, one may keep multiple terms on
the right hand side if the value of $\sum_I n_I\Delta_I$ is the same.
In this case, the moduli are partly unlifted. If we can interpret the integration
over the unlifted moduli space as the 1-loop zero-mode integral in the large $N$
calculation, only $\sum_I n_I\Delta_I$ would appear in the leading
large $N$ free energy since this is the only invariant quantity on the
unlifted moduli space. Of course this line of thinking assumes many things, such
as the equivariant localization picture of the index, etc. However, we think
it is a somewhat natural explanation of the appearance of $\sum_I n_I\Delta_I$.

\cite{Arai:2020uwd,Gaiotto:2021xce,Imamura:2022aua} explored the index of the 6d
SCFTs on $N$ M5-branes from M2-brane giant gravitons,
and also the index of the 3d SCFTs on $N$ M2-branes from M5-brane giant
gravitons. We shall now study the 6d/3d indices accepting the existence
of such giant graviton expansions, only assuming the analytic continuation picture
and the $n_I$ dependence through $\sum_I n_I\Delta_I$. In the former case,
the M2-brane giant gravitons wrap an internal $S^2\subset S^4$. So analytic
continuations of the 3d maximal SCFT index will give the giant graviton index in
$AdS_7\times S^4$. In the latter case, the M5-brane giant gravitons wrap an
internal $S^5\subset S^7$. So the 6d maximal SCFT index will provide
the giant graviton index in $AdS_4\times S^7$. Integrating out
the giant graviton numbers, we indeed recover the free energies and entropies
of $AdS_{7,4}$ black holes.

\hspace*{-.65cm}{\bf \underline{$AdS_4$ black holes from M5-branes}:}
We assume the following giant graviton expansion \cite{Arai:2020uwd} of the
3d index of maximal SCFT living on $N$ M2-branes \cite{Bhattacharya:2008zy}:
\begin{equation}
  Z_{\rm 3d}(\Delta_{1,2,3,4};\omega)=Z_{\rm KK}
  \sum_{n_1,n_2,n_3,n_4=0}^\infty e^{-N\sum_{I=1}^4\Delta_I n_I}
  Z_{n_1,n_2,n_3,n_4}(\Delta_I;\omega)\ .
\end{equation}
$\Delta_I$ are for the $U(1)^4\subset SO(8)$ R-symmetry
and $\omega$ is for the $U(1)\subset SO(3)$ rotation, satisfying $\sum_I\Delta_I-\omega=2\pi i\mathbb{Z}$.
$n_I$ are the numbers of maximal giant gravitons wrapping four
$S^5\subset S^7$.

When only one $n_I$ is nonzero, say when $n_1\neq 0$, $\log Z_{n_1,0,0,0}$ is obtain
by analytically continuing the free energy $\log Z^{\rm 6d}_{n_1}$ of 6d $(2,0)$ SCFT
of $A_{n_1-1}$ type at large $n_1$.
The three spacetime parameters are $\Delta_{2,3,4}$, and the two
$U(1)^2$ internal parameters are $\omega,-\Delta_1$.
Either from QFT \cite{Choi:2018hmj,Nahmgoong:2019hko} or gravity \cite{Hosseini:2018dob}
considerations, one obtains
\begin{equation}\label{M5-giant-single}
  \log Z_{n_1,0,0,0}=\log Z^{\rm 6d}_{n_1}
  \equiv -\frac{n_1^3}{24}\frac{\omega^2(-\Delta_1)^2}{\Delta_2\Delta_3\Delta_4}
\end{equation}
at $\sum_I\Delta_I-\omega=\pm 2\pi i$.
When all $n_I$'s are nonzero, there would be four different 6d QFT's, and
also extra modes supported on the intersection $S^3$ of two giants and
on the intersection $S^1$ of three giants. We suggest that the
net large $n_I$ free energy is given by
\begin{equation}\label{M5-conjecture}
  \log Z_{n_1,n_2,n_3,n_4}= -\frac{\omega^2\left(\sum_{I=1}^4 n_I\Delta_I\right)^3}
  {24\Delta_1\Delta_2\Delta_3\Delta_4}\ .
\end{equation}
This formula reduces to the expected ones like (\ref{M5-giant-single})
when only one $n_I$ is nonzero. This is the unique expression with the correct
limits and the $\sum_I n_I\Delta_I$ dependence.

Expanding the numerator of (\ref{M5-conjecture}),
one obtains contributions from the 6d/4d/2d modes:
\begin{equation}
  \log Z_{n_1,n_2,n_3,n_4}= \sum_{I=1}^3\log Z^{\rm 6d}_{n_I}
  +\sum_{I<J}\log Z^{\rm 4d}_{n_I,n_J}+\sum_{I<J<K}\log Z^{\rm 2d}_{n_I,n_J,n_K}\ .
\end{equation}
$\log Z^{\rm 6d}_{n_I}$ are given by (\ref{M5-giant-single}) or its permuted versions,
and the other terms are given by
\begin{equation}\label{free-intersection}
  \log Z^{\rm 4d}_{n_1,n_2}= -\frac{n_1n_2\omega^2
  \left(n_1\Delta_1+n_2\Delta_2\right)}{8\Delta_3\Delta_4}\ \ ,\ \
  \log Z^{\rm 2d}_{n_1,n_2,n_3}=
  -\frac{n_1n_2n_3\omega^2}{4\Delta_4}\ .
\end{equation}
We can independently justify them.  
$\log Z^{\rm 4d}_{n_1,n_2}$ is the free energy of an SCFT at the intersection of
$n_2$ M5-branes on $012345$, and $n_1$ M5-branes on $01236,10$. Its free energy takes
the form of $\frac{P(-\Delta_1,-\Delta_2,\omega)}{\Delta_3\Delta_4}$, where $P$ is the cubic anomaly polynomial for $SO(2)_{45}$, $SO(2)_{6,10}$,
$SO(3)_{789}$.\footnote{$-\Delta_1$, $-\Delta_2$ are inserted since they correspond to
the tachyonic transverse directions. } The $SO(2)_{45}$-$SO(3)$-$SO(3)$ anomaly can be computed by separating the $n_2$ M5-branes along the $6$'th direction, and compactifying the $10$'th direction to obtain Witten's $SU(n_1)^{n_2-1}$ MQCD \cite{Witten:1997sc}.
($SO(2)_{6,10}$ is explicitly broken by the deformations.)
This is a linear quiver with $n_1$ fundamentals attached to each $SU(n_1)$ node at
the end. The large $n_I$ anomaly is given by
\begin{equation}
  k_{\Delta_1\omega\omega}\equiv
  {\rm Tr}\left[J_{45}J_{78}J_{78}\right]=\frac{n_{\rm V}}{4}
  =\frac{(n_1^2-1)(n_2-1)}{4}\approx\frac{n_1^2n_2}{4}\ ,
\end{equation}
where $n_{\rm V}$ is the number of vector multiplets. This yields the
following contribution to the anomaly polynomial $P$ (e.g. see eqn.(2.34) of \cite{Kim:2019yrz}):
\begin{equation}
  \frac{3k_{\Delta_1\omega\omega}(-\Delta_1)\omega^2}{6}
  =-\frac{n_1^2n_2\Delta_1\omega^2}{8}\ .
\end{equation}
This explains the first term of $\log Z^{\rm 4d}_{n_1,n_2}$ in (\ref{free-intersection}).
Similarly, its second term is explained from the $SO(2)_{6,10}$-$SO(3)^2$
anomaly of the $SU(n_2)^{n_1-1}$ MQCD. $\log Z^{\rm 2d}_{n_1,n_2,n_3}$ can also be
computed from anomalies, but here we just explain a quick check of its coefficient
from the entropy of 4d black holes obtained by triply intersecting M5-strings
with momentum $p$. The entropy is
given by $2\pi\sqrt{n_1n_2n_3p}$, which is obtained by extremizing
$\frac{\pi^2n_1n_2n_3}{\beta}+p\beta$. This is the Cardy limit of $\log Z_{n_1,n_2,n_3}^{\rm 2d}$ in (\ref{free-intersection}),
upon taking $\Delta_4=\beta\ll 1$ and $\omega=\sum_I\Delta_I\mp 2\pi i\approx\mp 2\pi i$.

Now we extremize the entropy function given by
\begin{equation}\label{AdS7-entropy-ftn}
  S(Q_I,J;\Delta_I,\omega,n_I)=\log Z_{n_1,n_2,n_3,n_4}
  +\sum_{I=1}^4\Delta_I(Q_I-Nn_I)+\omega J
\end{equation}
with (\ref{M5-conjecture}). As in section 2, we may either understand it as
maximizing ${\rm Re}[S]$ with real $n_I$'s for small black holes, or extremizing
$S$ with complex $n_I$'s for generic black holes. We present the calculation in
the latter viewpoint. Extremizing $S$ in $n_I$, $\sum_I n_I\Delta_I$ is given by
\begin{equation}\label{AdS7-roots}
  \left(\sum_I n_I\Delta_I\right)^2=-\frac{8N\Delta_1\Delta_2\Delta_3\Delta_4}
  {\omega^2}\ \rightarrow\
  \sum_I n_I\Delta_I=\pm 2\sqrt{2}i N^{\frac{1}{2}}
  \frac{\sqrt{\Delta_1\Delta_2\Delta_3\Delta_4}}{\omega}\ .
\end{equation}
We should pick one saddle point solution.
To explain this, we employ the convention of \cite{Choi:2018fdc,Choi:2019zpz} on
the square-root, which sets $\sqrt{\Delta^4}=-\Delta^2$ when
all $\Delta_I$'s are equal. Then, among the two solutions of (\ref{AdS7-roots}),
one should choose the upper/lower sign at $\sum_{I=1}^4\Delta_I-\omega=\pm 2\pi i$,
respectively. We showed this by a Picard-Lefschetz analysis like that of section 2.2.
With this choice, let us first discuss the small black hole limit,
in which case $\Delta_I\ll 1$ are small positive numbers and
$\omega\approx \mp 2\pi i$. Then (\ref{AdS7-roots}) at equal $\Delta_I$'s reduces to
\begin{equation}
  \Delta\sum_I n_I\approx\pm 2\sqrt{2}iN^{\frac{1}{2}}\cdot\frac{-\Delta^2}{\mp 2\pi i}
  =2\sqrt{2}N^{\frac{3}{2}}\Delta^2\ ,
\end{equation}
yielding real positive $\sum_I n_I$. This ensures that extremizing $S$ with complex $n_I$
is equivalent to maximizing ${\rm Re}[S]$ with real $n_I$ in the small black hole
limit.

Inserting the solution picked in the previous paragraph
into (\ref{AdS7-entropy-ftn}),  one obtains
\begin{equation}
  S(Q_I,J;\Delta_I,\omega)=\mp \frac{4\sqrt{2}iN^{\frac{3}{2}}}{3}
  \frac{\sqrt{\Delta_1\Delta_2\Delta_3\Delta_4}}{\omega}
  +\sum_{I=1}^4\Delta_I Q_I+\omega J
\end{equation}
which perfectly reproduces the entropy function of BPS black holes
in $AdS_4\times S^7$ \cite{Choi:2018fdc}.

\hspace*{-.65cm}{\bf \underline{$AdS_7$ black holes from M2-branes}:}
We assume the following giant graviton expansion \cite{Arai:2020uwd}
of the index of 6d $(2,0)$ SCFT of $A_{N-1}$ type \cite{Bhattacharya:2008zy}:
\begin{equation}
  Z_{\rm 6d}(\Delta_1,\Delta_2;\omega_1,\omega_2,\omega_3)=Z_{\rm KK}
  \sum_{n_1,n_2=0}^\infty e^{-N\sum_{I=1}^2\Delta_I n_I}Z_{n_1,n_2}
  (\Delta_I;\omega_i)\ .
\end{equation}
Here $\Delta_{1,2}$ are for the $U(1)^2\subset SO(5)$
R-symmetry, and $\omega_{1,2,3}$ are for the $U(1)^3\subset SO(6)$ in the spacetime.
They satisfy $\sum_I\Delta_I-\sum_i\omega_i=2\pi i\mathbb{Z}$.
$n_I$ are the numbers of maximal giant gravitons wrapping two different
$S^2\subset S^4$ cycles.

Assuming the analytic continuation picture, $\log Z_{n_1,0}$ at large $n_1\sim N$
will be given by the 3d free energies on $n_1$ M2-branes.
Taking $\omega_{1,2,3},-\Delta_1$ as the internal parameters and $\Delta_2$ as
the worldvolume parameter, the analytically continued free energy is
given by \cite{Choi:2018fdc,Choi:2019zpz,Nian:2019pxj,Choi:2019dfu}
\begin{equation}
  \pm\frac{4\sqrt{2}in_1^{\frac{3}{2}}}{3}
  \frac{\sqrt{\omega_1\omega_2\omega_3(-\Delta_1)}}{\Delta_2}\ \ ,\ \
  \sum_{I=1}^2\Delta_I-\sum_{i=1}^3\omega_i=\pm 2\pi i\ .
\end{equation}
Similar expression is
obtained for $\log Z_{0,n_2}$.
Now we suggest that the general $\log Z_{n_1,n_2}$ at large $n_I\sim N$ is
given by
\begin{equation}\label{M2-conjecture}
  \log Z_{n_1,n_2}\sim\pm\frac{4\sqrt{2}i}{3}
  \sqrt{\omega_1\omega_2\omega_3}
  \frac{\left(-n_1\Delta_1-n_2\Delta_2\right)^{\frac{3}{2}}}{-\Delta_1\Delta_2}\ .
\end{equation}
This gives the desired limits when either of $n_1,n_2$ vanishes,
and depends only on $\sum_I n_I\Delta_I$. It defies factorization between 3d-1d
degrees of freedom.

Now extremizing the entropy function given by
\begin{equation}\label{entropy-ftn-AdS7}
  S(Q_I,J_i;\Delta_I,\omega_i,n_I)=\log Z_{n_1,n_2}+
  \sum_{I=1}^2\Delta_I(Q_I-Nn_I)+\sum_{i=1}^3\omega_i J_i
\end{equation}
with (\ref{M2-conjecture}), one obtains
\begin{equation}
  \pm 2\sqrt{2}i\sqrt{\omega_1\omega_2\omega_3(-n_1\Delta_1-n_2\Delta_2)}
  =N\Delta_1\Delta_2
  \ \rightarrow\ \sum_{I=1}^2n_I\Delta_I=\frac{N^2}{8}\frac{(\Delta_1\Delta_2)^2}
  {\omega_1\omega_2\omega_3}\ .
\end{equation}
Inserting this back to the entropy function, one obtains
\begin{equation}
  S(Q_I,J_i;\Delta_I,\omega_i)=-\frac{N^3}{24}\frac{\Delta_1^2\Delta_2^2}
  {\omega_1\omega_2\omega_3}+\sum_{I=2}^2\Delta_I Q_I+\sum_{i=1}^3\omega_i J_i
\end{equation}
which is precisely the entropy function of BPS black holes in
$AdS_7\times S^4$ \cite{Hosseini:2018dob}.

Before finishing this exercise on $AdS_7\times S^4$, we comment on
a puzzle that we resolved only partly. In section 2, it was important to have a
leading order degeneracy after fixing $\sum_I n_I\Delta_I$ by extremization, for
the continuum conjecture and cancellations of $Z_{n_I}$'s.
Here, fixing $n_1\Delta_1+n_2\Delta_2$ at general complex $\Delta_I$'s does not
leave any degeneracy. So one may wonder whether the picture of section 2 is valid
here. Curiously, we can see that the apparent real maximum $n_{I}^0$ of
${\rm Re}[S(Q_I,J_i,n_I)]$ is generally subject to severe cancellations already with
the leading free energy. To simplify the discussions, let us unrefine
$\Delta_I$, $\omega_i$ and keep $n\equiv n_1+n_2$,
$q\equiv \frac{Q_1+Q_2}{2}+\frac{1}{3}\sum_iJ_i$ only. Then the entropy scales
like
\begin{equation}
  S(q,n)=N^3f({\textstyle \frac{q}{N^3}, \frac{n}{N^2} })\ .
\end{equation}
The maximum $n_0$ satisfies ${\rm Re}[f^\prime(\frac{n_0}{N^2})]=0$. Now considering
a neighborhood of $n_0$, one finds
\begin{equation}
  S\left({\textstyle \frac{n_0+\Delta n}{N^2}}\right)=
  N^3f\left({\textstyle \frac{n_0}{N^2}}\right)+
  Nf^\prime\left({\textstyle \frac{n_0}{N^2}}\right)\Delta n+
  \frac{1}{2N}f^{\prime\prime}\left({\textstyle \frac{n_0}{N^2}}\right)(\Delta n)^2
  +\cdots\ .
\end{equation}
Since the real part of the second term vanishes at $n_0$, the change of
${\rm Re}[S]$ is very slow in a wide range of  $|\Delta n|$ ($\ll N^{\frac{1}{2}}$).
However, ${\rm Im}[S]$ generally varies fast in this neighborhood due to
the second term $N{\rm Im}[f^\prime(\frac{n_0}{N^2})]\Delta n$. This causes
cancellations of nearby terms around $n=n_0$, reducing the
apparently over-estimated entropy at $n_0$. This cancellation happens because
of the large $N$ scalings $S\sim N^3$ and $n\sim N^2$: similar cancellations do not
happen in AdS$_5$/CFT$_4$ with $S\sim N^2$ and $n\sim N$. This
provides the expected cancellations of the giant graviton index. We would like to
further understand what makes the full continuum conjecture for $n_I$'s possible.

\section{Conclusion}

In this paper, we studied the microstate counting of
BPS black holes in AdS$_{4,5,7}$ from the viewpoint of giant graviton expansions.

We employed a saddle point approach to compute the large $N$ giant graviton
index $Z_{n_1,n_2,n_3}$ in $AdS_5\times S^5$, when the three electric chemical
potentials are unrefined.
Further maximizing the corresponding entropy in $n_I$, we successfully accounted
for the microstates of small AdS$_5$ black holes. To understand the black holes
at general sizes, one had to take into account the cancellations of
$Z_{n_1,n_2,n_3}$'s. We conjectured a particular cancellation
mechanism which successfully reproduces the
entropies of the black holes at arbitrary sizes. We further proposed the large $N$
free energy for $Z_{n_1,n_2,n_3}$ with all chemical potentials refined.

We interpreted the large $N$ giant graviton free
energies as the analytically continued indices of SCFT's. Extending this
interpretation to other AdS/CFT examples, we found intriguing connections between
the SCFT free energies in various dimensions. We obtained the large
$N$ free energies of 6d/3d SCFTs on M5/M2-branes from the analytic continuations
of vice versa, after suitably dressing them by certain defect free energies.
This suggests that the (BPS) spectrum of quantum gravity
admits dual formulations, either in terms of electric or magnetic variables.
Presumably, there will be more duality relations of this sort that one can find from
giant gravitons and analytic continuation. As a crude but nontrivial exercise,
we also tried to account for the large $N$ free energy of the 5d SCFTs dual
to the massive IIA theory on $AdS_6\times S^4/\mathbb{Z}_2$ \cite{Choi:2018fdc,Choi:2019miv,Crichigno:2020ouj}, from analytic
continuations of the free energy of D2-brane SCFTs on an orbifold.
Although the details need to be clarified, we find that the $N^{\frac{5}{2}}$
scaling of the former free energy is obtained from the
$n^{\frac{5}{3}}$ scaling for D2-branes in massive IIA theory.

Perhaps we should also comment on our current understanding of the giant graviton
expansion. The formula was constructed from maximal giant gravitons, wrapping
the largest (non-topological) $S^3\subset S^5$. The worldvolume degrees of
freedom include tachyons which shrink the giant gravitons.
The tachyonic quiver theories on these branes were
used in a subtle way \cite{Imamura:2021ytr} to write down the
giant graviton index formula. Technically, the tachyonic part of the
partition function was first analytically continued. Then one chose the path
integral contour for the holonomy zero modes empirically. After these ad hoc steps,
the resulting $Z_{n_1,n_2,n_3}$ appears to correctly describe the spectral problem
of the Yang-Mills index. However, its physics looks different from that of the
original quiver system. Namely, the `ground states' of $Z_{n_1,n_2,n_3}$
describe the trace relation constraints, subtracting the over-estimated states in the graviton index. We feel that a well-defined QFT cannot show this behavior.
So it seems desirable to better understand the relation between
the quiver and the formula.

Giant gravitons were originally conceived in \cite{McGreevy:2000cw} as
a gravity mechanism of imposing trace relation constraints at $E\sim N^1$.
However, in this paper, we found that their worldvolume degrees of freedom
are also responsible for enhanced entropies at $E\sim N^2$.
Perhaps new BPS states exist thanks to the trace relations.
Since crucial roles were played by the open strings connecting various
D-branes in our studies, one can construct an ansatz for the BPS
cohomologies in terms of open spin chains ending on determinant operators
\cite{Hofman:2006xt}.
We expect that such  an ansatz will be relevant at least in the small black
hole regime, $q\ll N^2$.

In section 2.2, we conjectured a possible mechanism in which various
$Z_{n_1,n_2,n_3}$'s are partly canceled to account for the black hole entropies
correctly. A key feature was the discrete sums of $n_I$'s being
effectively replaced by integrals. We have no a priori justification for
this conjecture, except that the final entropy derived with this assumption
is correct. Here we simply want to note that the discrete points labeled by
$(n_1,n_2,n_3)$ are special points on the moduli space of $\frac{1}{8}$-BPS
giant gravitons \cite{Mikhailov:2000ya}. As alluded to in
\cite{Gaiotto:2021xce,imamura-talk},
they could be the fixed points of the equivariant localization calculation
of an integral over this moduli space. Perhaps it may be helpful to seek 
such an integral reformulation of this sum to justify our continuum conjecture.

\vskip 0.5cm

\hspace*{-0.8cm} {\bf\large Acknowledgements}
\vskip 0.2cm

\hspace*{-0.75cm} We thank Hee-Cheol Kim, Kimyeong Lee, Jaemo Park and Jaewon Song
for helpful discussions. This work is supported in part by a KIAS Individual Grant
(PG081601) at Korea Institute for Advanced Study (SC) and
the National Research Foundation of Korea (NRF) Grant 2021R1A2C2012350 (SK, EL, JL).

\end{document}